\documentclass[a4paper,12pt]{article}
\usepackage{marvosym}
\usepackage{comment}
\usepackage{subfigure}
\usepackage{mathrsfs} 
\usepackage[framemethod=default]{mdframed}

\newmdenv[skipabove=1pt,
skipbelow=1pt,
rightline=false,
leftline=false,
topline=false,
bottomline=false,
backgroundcolor=gray!10,
linecolor=gray,
innerleftmargin=5pt,
innerrightmargin=5pt,
innertopmargin=-10pt,
innerbottommargin=5pt,
leftmargin=0cm,
rightmargin=0cm,
linewidth=4pt]{eBox}
\catcode`\,12

\pdfoutput=1 % if your are submitting a pdflatex (i.e. if you have
\usepackage{jheppub} % for details on the use of the package, please
\usepackage[T1]{fontenc} % if needed

\title{Phases of String Stars in the Presence of a Spatial Circle}

\author{Jinwei Chu}

\affiliation{Department of Physics, University of Chicago, Chicago, IL 60637, USA}

\emailAdd{jinweichu@uchicago.edu}
\abstract{In string theory, black holes are expected to transition into string stars as their Hawking temperatures approach the Hagedorn temperature. We study string stars and their phase transitions in the Euclidean spacetime $\mathbb{R}^d\times\mathbb{S}_\tau^1\times\mathbb{S}_z^1$. Using the Horowitz-Polchinski (HP) effective field theory, we discover novel solutions for $d=2$. The uniform string star exhibits a scaling symmetry that results in the absence of a critical point for its transition into the non-uniform solution. For $d=4$, we show that quartic corrections to the effective action resolve the mass degeneracy of uniform string stars. At $d=5$, we find that as non-uniformity increases, the quartic terms become significant (while higher-order terms remain negligible) and reverse the direction of temperature variation, leading to a swallowtail-type phase diagram in the canonical ensemble. Extending the quartic-corrected EFT to $d=6$, we find that string stars with small non-uniformity dominate the microcanonical ensemble but not the canonical ensemble, similar to the $d=5$ case. However, in the microcanonical ensemble, the uniform string star is anomalously (un)stable when the spatial circle is larger (smaller) than the critical size.
}

\usepackage{amsmath,amssymb,amsfonts,mathtools,bm} % Default math packages
\numberwithin{equation}{section}
\usepackage{physics} % Default physics package
\usepackage[margin=10pt,font=small,labelfont=bf]{caption}
\usepackage[pdftex,dvipsnames]{xcolor}
\definecolor{darkblue}{rgb}{0.1,0.1,.7}
\usepackage{graphicx}  % Allows including images
\graphicspath{ {images/} }
\mathcode`l="8000
\begingroup
\makeatletter
\lccode`\~=`\l
\DeclareMathSymbol{\lsb@l}{\mathalpha}{letters}{`l}
\lowercase{\gdef~{\ifnum\the\mathgroup=\m@ne \ell \else \lsb@l \fi}}%
\endgroup
\def\<#1\>{\expval{#1}}

\newcommand{\undertl}[1]{\mathord{\vtop{\ialign{##\crcr
				$\hfil\displaystyle{#1}\hfil$\crcr\noalign{\kern1.5pt\nointerlineskip}
				$\hfil\tilde{}\hfil$\crcr\noalign{\kern1.5pt}}}}} % Puts tilde under given variable

\newcommand   \lptl{\raise .8ex\hbox{$^\leftarrow$} \hspace{-9pt} \partial} % derivatives with arrows
\newcommand   \lrptl{\raise .8ex\hbox{$^\leftrightarrow$} \hspace{-9pt} \partial} % derivatives with arrows
\def\be#1\ee{\begin{equation}\begin{aligned}#1\end{aligned}\end{equation}}
\makeatletter
\DeclareRobustCommand\bea{\@ifnextchar[{\@@bea}{\@bea}}
\def\@@bea[#1]#2\eea{\begin{subequations}\begin{align}#2\end{align}\label{#1}\end{subequations}}
\def\@bea#1\eea{\begin{subequations}\begin{align}#1\end{align}\end{subequations}}
\makeatother
 \catcode`,\active
 
 \catcode`\,12
 
\renewcommand \l  {\lambda}

\begin{document} 
\maketitle
\flushbottom
\section{Introduction and summary}
A Schwarzschild black hole is a nontrivial vacuum solution of general relativity (GR) that exhibits spherical symmetry. The black hole features an event horizon with a radius of $r_H$ and a singularity at the center. Near the singularity, curvatures blow up, leading to the breakdown of GR. Due to this, the Schwarzschild black hole remains incompletely understood. 

In the Euclidean signature, where the spacetime becomes $\mathbb{R}^d\times \mathbb{S}_\tau^1$ via a Wick rotation $\tau =it$, the black hole solution is under good control. In particular, the Euclidean time circle $\mathbb{S}_\tau^1$ is capped off at the horizon radius $r_H$, effectively removing the singularity. This can be seen from the metric,
\begin{equation}
\label{schw}
    ds^2=\left(1-\left(\frac{r_H}{r}\right)^{d-2}\right)d\tau^2+\frac{dr^2}{1-\left(\frac{r_H}{r}\right)^{d-2}}+r^2d\Omega_{d-1}^2\ .
\end{equation}

In (\ref{schw}), the circumference of the $\tau$ circle corresponds to the inverse Hawking temperature $\beta$, which is related to the horizon radius by
\begin{equation}
    r_H=\frac{(d-2)\beta}{4\pi}\ .
\end{equation}
This relation also lets $\mathbb{S}_\tau^1$ smoothly capped off at $r_H$.

Embedded in string theory, the Schwarzschild metric (\ref{schw}) receives string corrections~\cite{CALLAN1989673,Chen:2021qrz}. When the length scale of the black hole is much larger than the string length, i.e., $\beta\gg l_s\equiv \sqrt{\alpha'}$, the string corrections are suppressed. However, as $\beta$ approaches $l_s$, GR breaks down, making the black hole solution less well understood.

When $\beta\sim l_s$, it is widely believed that the black hole transitions into a state of highly excited fundamental strings~\cite{Bowick:1985af,Susskind:1993ws,Horowitz:1996nw} (see also \cite{Sen:1995in,Damour:1999aw,Khuri:1999ez,Kutasov:2005rr,Giveon:2005jv,Giveon:2006pr} for relevant works). Furthermore, near a string-scaled temperature known as the ``Hagedorn temperature'', denoted by $\beta_H$, there exists an effective field theory (EFT) proposed by Horowitz and Polchinski~\cite{Horowitz:1997jc}. Similar to GR, the Horowitz-Polchinski (HP) EFT admits classical solutions, commonly referred to as ``string stars''. In particular, the spherically symmetric HP solution is of interest as an analog to the Schwarzschild black hole. For recent developments, see~\cite{Brustein:2021cza,Chen:2021emg,Chen:2021dsw,Urbach:2022xzw,Balthazar:2022szl,Balthazar:2022hno,Bedroya:2022twb,Urbach:2023npi,Ceplak:2023afb,Halder:2023nlp,Agia:2023skp,Bedroya:2024uva,Santos:2024ycg,Chu:2024ggi,Emparan:2024mbp,Ceplak:2024dxm,Bedroya:2024igb}.

So far, we have assumed that the space is non-compact, i.e., $\mathbb{R}^d$. However, to explain certain physical phenomena, it is often useful to include compact spatial dimensions as well (e.g., the recent dark dimension scenario proposed within the Swampland program~\cite{Montero:2022prj}). Therefore, understanding the solutions in the presence of compact dimensions becomes crucial. In this paper, we consider an additional spatial circle $\mathbb{S}_z^1$ with periodicity $z\sim z+L$, focusing on the case where $L\gg l_s$. 

The introduction of an extra spatial circle gives rise to new solutions, especially those that are non-uniform along $\mathbb{S}_z^1$. Consequently, there is a competition between uniform and non-uniform solutions in the canonical and microcanonical ensembles. In the case of black holes, where the Schwarzschild metric (\ref{schw}) is uplifted to a black string uniformly wrapped around the $z$ circle, the phase diagram and transition between uniform and non-uniform solutions have been extensively studied (see, for example,~\cite{Gubser:2001ac,Wiseman:2002zc,Kol:2002xz,Sorkin:2004qq,Kol:2004pn,Kudoh:2005hf,Kol:2006vu,Harmark:2007md,Asnin:2007rw,Emparan:2015gva,Emparan:2018bmi,Figueras:2022zkg}). Notably, in Lorentzian spacetime, it was discovered by Gregory and Laflamme that the uniform black string becomes unstable to non-uniform perturbations when the spatial circle is large compared to the horizon radius~\cite{Gregory:1993vy,Gregory:1994bj}.

Unlike the black holes, the Horowitz-Polchinski (HP) solutions have no natural extensions in Lorentzian signature, making it ambiguous to discuss their Gregory-Laflamme instability. However, in Euclidean spacetime, we can analogously examine the instability of the free energy or entropy~\cite{Urbach:2022xzw}. Specifically, by finding the time-independent non-uniform perturbation at the leading order, the critical point of the instability towards non-uniformity can be determined. Furthermore, the phase diagram can be constructed by comparing the free energy or the entropy of the uniform and non-uniform solutions.

In~\cite{Chu:2024ggi}, we use the first-order non-uniform perturbation to determine the critical point of instability for the HP solution towards non-uniformity. By analyzing the second-order non-uniform perturbation, we further determine the order of phase transition at the critical point. For example, it is shown that for $d>4$ ($d\le 4$), in the canonical ensemble, the uniform HP solution undergoes a first (second)-order phase transition at the critical temperature and becomes non-uniform.
On the other hand, in the microcanonical ensemble, for $d<4$ ($d\ge 4$), the uniform HP solution undergoes a first (second)-order phase transition at the critical mass. We also sketch the phase diagram for the microcanonical ensemble there. Subsequently,~\cite{Emparan:2024mbp} obtained the non-uniform HP solutions beyond perturbative analysis and presented the phase diagrams of both the canonical and microcanonical ensembles for $d=3$, 4 and 5.

In this paper, we continue the study of string star solutions in the presence of the spatial circle $\mathbb{S}_z^1$. In section \ref{secHP}, we analyze the Horowitz-Polchinski EFT. For $d=2$, where the standard HP solution ceases to exist, we find novel solutions in which the Euclidean time circle does not asymptotically converge to a constant, namely $\beta$. Although $\beta$ no longer has a physical interpretation, such as representing the inverse temperature, it serves as a parameter for the non-uniform solutions. In contrast, the uniform solutions are not determined by $\beta$ due to a scaling transformation that relates different solutions. As a result, the uniform and non-uniform solutions are not connected with each other.

When we treat $d$ as a continuous parameter and increase it slightly above 2, standard HP solutions emerge in which the $\tau$ circle converges to a length of $\beta$. In this case, we start with a localized higher-dimensional HP solution and gradually decrease $\beta$. We find that the non-uniform solution eventually transitions into a uniform solution at a critical temperature. Phase diagrams of both the canonical and microcanonical ensembles are presented for $d$ between 2 and 3. In the canonical ensemble, the non-uniform phase connects continuously to the uniform phase at the critical point, indicating a second-order phase transition. This finding is consistent with the perturbative analysis in~\cite{Chu:2024ggi}.

In the microcanonical ensemble, there are two branches of non-uniform HP solutions that meet at a mass distinct from the critical point. The lower non-uniform branch extends to the critical mass, where it intersects with the uniform branch. Together, the uniform and non-uniform solutions form a swallowtail-type phase diagram, indicating a first-order phase transition consistent with~\cite{Chu:2024ggi}. 

For $d=3$, we reproduce the phase diagrams presented in~\cite{Emparan:2024mbp}. In the canonical ensemble, as $\beta$ decreases, the non-uniform HP solution undergoes a second-order phase transition into a uniform HP solution, similar to the behavior observed for $2<d<3$. However, in the microcanonical ensemble, the uniform solution consistently has a higher entropy than the non-uniform solution, which contrasts with the case of $2<d<3$. Consequently, the uniform HP solution does not transition into a non-uniform HP solution as the mass varies. 

Nevertheless, this does not rule out the possibility of the HP solution becoming non-uniform at the critical mass. Indeed, as proposed in~\cite{Chu:2024ggi}, it is likely to undergo a first-order phase transition into a localized black hole at this point.

Similarly, for $d=4$, the free energy of the non-uniform HP solution changes continuously to that of the uniform phase at the critical temperature. However, in the microcanonical ensemble, the mass of the uniform HP solution remains fixed as $\beta$ varies. To resolve this mass degeneracy, we incorporate the quartic terms in the effective action\footnote{I thank David Kutasov for pointing this out.}, as shown in section \ref{secHP4}. As a result, we find that the mass of the uniform string star increases with $\beta$.

In section \ref{secHP4}, we also use the quartic-corrected effective action to study string stars in $\mathbb{R}^5\times \mathbb{S}_z^1$. For the case of noncompact spatial dimensions, i.e., $\mathbb{R}^d$, it is known that the original HP EFT breaks down as $d$ approaches 6 when treated as a continuous parameter. This can be demonstrated through a scaling analysis~\cite{Balthazar:2022hno} (see also \cite{Chu:2024ggi} for another derivation based on the free energy). However, the EFT can be rescued by merely incorporating the quartic terms into the HP effective action. These suggest that in $\mathbb{R}^5\times \mathbb{S}_z^1$, as the string star becomes increasingly non-uniform and approaches a localized higher-dimensional solution, the quartic terms become significant and must be taken into account.

By analyzing the quartic-corrected EFT, we find that for $d=5$, the transition between the uniform and non-uniform phases is first-order in the canonical ensemble, whereas it is second-order in the microcanonical ensemble, contrasting with the behavior observed for $d\le 4$. These results are also consistent with the perturbative analysis in~\cite{Chu:2024ggi}. Notably, if the quartic terms are neglected, the free energy of the non-uniform phase is always higher than that of the uniform phase~\cite{Emparan:2024mbp}, leading to a puzzle regarding the end state of the phase transition from the uniform phase at the critical point. Our findings show that when the quartic terms are included, an additional branch of non-uniform solutions emerges, with a lower free energy than the uniform phase at the critical point, thereby resolving the puzzle.

Finally, in $\mathbb{R}^{d>6}$, even the quartic-corrected effective action becomes invalid, because the field amplitudes are not small at any temperature, which means that higher-order terms are no longer suppressed~\cite{Balthazar:2022hno}. To gain further insights into higher-dimensional string stars, we study the quartic-corrected EFT in $\mathbb{R}^6\times \mathbb{S}_z^1$ and examine the behavior of the solution as its non-uniformity increases. Specifically, we begin with the uniform solution and introduce non-uniform perturbations, allowing us to generate the entire non-uniform branch. 

In the canonical ensemble, we find that the non-uniform branch emerges from a critical point on the uniform branch. Near this critical point, the non-uniform solutions correspond to inverse temperatures below the critical value and possess higher free energies than the uniform solutions. This means that below the critical inverse temperature, the uniform phase is stable against non-uniformity, and the phase transition at the critical temperature is first-order. Note that when the non-uniformity is small, the EFT is expected to remain valid.

As the non-uniformity increases, the free energy of the solution remains higher than that of the uniform solution. This suggests that the uniform string star never transitions into the non-uniform one. However, we also find a distinct non-uniform branch that extends to the Hagedorn temperature and has much lower free energy than the uniform solution. Therefore, in $\mathbb{R}^6\times\mathbb{S}^1$, the uniform string star indeed undergoes a first-order phase transition into a non-uniform one at the critical temperature, which is the same as the case of $d=5$.

The behavior of the microcanonical ensemble in $\mathbb{R}^6\times \mathbb{S}^1$ is more unusual. Although the phase transition is second-order, as in the $d=5$ case, we unexpectedly find that the uniform solutions stable in the canonical ensemble become unstable towards non-uniformity, as they coexist with the non-uniform branch possessing higher entropy. On the other hand, the uniform solutions that are unstable in the canonical ensemble are now stable. This phenomenon does not occur in any of the other cases we study.
\section{Horowitz-Polchinski EFT in $\mathbb{R}^d\times\mathbb{S}_z^1$}\label{secHP}
First, consider a $(d+1)$-dimensional noncompact spacetime $\mathbb{R}^{d,1}$. To study the canonical ensemble, we perform a Wick rotation of the time coordinate, $t\to\tau=it$, resulting in the Euclidean spacetime $\mathbb{R}^d\times\mathbb{S}_\tau^1$. The circumference of the Euclidean time circle $\mathbb{S}_\tau^1$ is identified with the inverse temperature $\beta$ of the canonical ensemble. 

In string theory, when the inverse temperature $\beta$ is near a critical value known as the Hagedorn temperature, $\beta_H$, the canonical ensemble can be described by the Horowitz-Polchinski (HP) effective field theory (EFT)~\cite{Horowitz:1997jc}. The HP EFT includes massless fields, such as the graviton, as well as the field denoted by $\chi$, which corresponds to the closed string tachyon with winding number $\pm 1$. The inclusion of $\chi$ arises because this winding tachyon, with a mass of $(\beta^2-\beta^2_H)/(2\pi\alpha')^2$, becomes light near $\beta_H$.

When all fields are small, the coupling between the field $\chi$ and the massless fields is dominated by its interaction with the radion $\varphi$, which captures the local variation of the Euclidean time circle. Specifically, at a given location $x$, the circumference of the $\tau$ circle is expressed as
\begin{equation}
    \beta(x)=\beta e^{\varphi(x)}\ .
\end{equation}
This leads to the effective mass of $\chi$ as follows
\begin{equation}
\label{meff}
    m^2_\text{eff}(x)=\frac{\beta^2e^{2\varphi(x)}-\beta^2_H}{(2\pi\alpha')^2}\ ,
\end{equation}
which establishes the coupling between the fields $\chi$ and $\varphi$.

Reducing on the $\tau$ circle, the EFT yields the following action in the flat space $\mathbb{R}^d$.
\begin{equation}
	\label{Sphichid}
	\begin{split}
 	I_d=\frac{\beta}{16\pi G_N^{(d+1)}}\int d^dx\left[(\nabla\varphi)^2+|\nabla \chi|^2+\left(m_\infty^2+\frac{\kappa}{\alpha'}\varphi\right)|\chi|^2\right]\ .
	\end{split}
	\end{equation}
Here, we truncate the action by neglecting higher-order terms and other fields that are either heavy or weakly coupled to $\chi$. Expanding (\ref{meff}) in $\varphi$ gives the mass term and coupling coefficient, i.e.,
\begin{equation}
\label{m2}
    m^2_\infty=\frac{\beta^2-\beta^2_H}{(2\pi\alpha')^2}\ ,\quad \kappa=\frac{\beta^2_H}{2\pi^2\alpha'}\ .
\end{equation}
The precise values of $\beta_H$ and $\kappa$ depend on the type of string theory. For example, in bosonic string theory, $\kappa=8$, while in type II string theory, $\kappa=4$.

Next, we introduce an additional spatial circle $\mathbb{S}_z^1$ with periodicity $z\sim z+L$. Assuming $L\gg l_s$, the effective action is given by the higher-dimensional analog of (\ref{Sphichid}),
\begin{equation}
\label{Sphichiz}
	\begin{split}
 	I_{d+1}=\frac{\beta}{16\pi G_N^{(d+2)}}\int_{-\frac{L}{2}}^{\frac{L}{2}}dz\int d^dx&\bigg[(\partial_z\varphi)^2+(\nabla_x\varphi)^2+|\partial_z \chi|^2+|\nabla_x \chi|^2\\
  &+\left(m_\infty^2+\frac{\kappa}{\alpha'}\varphi\right)|\chi|^2\bigg]\ .
	\end{split}
	\end{equation}
The $(d+2)$-dimensional Newton's constant, $G_N^{(d+2)}$, is related to its lower-dimensional counterpart by
\begin{equation}
    G_N^{(d+2)}=G_N^{(d+1)}L\ .
\end{equation} 

From the action (\ref{Sphichiz}), the e.o.m. for the fields $\chi$ and $\varphi$ are given by
 \begin{equation}
        \label{eomHP}
	\begin{split}
 	\nabla^2_x\chi+\partial_z^2\chi&=\left(m_{\infty}^2+\frac{\kappa}{\alpha'}\varphi\right)\chi \ ,\\
 	\nabla^2_x\varphi+\partial_z^2\varphi &=\frac{\kappa}{2\alpha'}|\chi|^2\ .
	\end{split}
	\end{equation}
In this paper, we focus on the cases with spherical symmetry and reflection symmetry. In other words, the fields $\chi$ and $\varphi$ depend only on the radial coordinate $r\equiv \sqrt{(x_1)^2+\cdots+(x_d)^2}$ and are symmetric under $z\to - z$. Additionally, we fix the field $\chi$ to be real.

To simplify the differential equations further, we apply the following rescaling:
\begin{equation}
\label{resHP}
	\begin{split}
	r&=\hat{r}/m_\infty\ ,\\
    z&=\hat{z}/m_\infty\ ,\\
 	\chi (r)&=\frac{\sqrt{2}\alpha' }{\kappa}m^2_{\infty}\hat{\chi}(\hat{r})\ ,\\
 	\varphi (r)&=\frac{\alpha'}{\kappa }m^2_{\infty}\hat{\varphi}(\hat{r})\ .
	\end{split}
	\end{equation}
Under this rescaling, the e.o.m. become
\begin{equation}
\label{eomHPres}
	\begin{split}
 	\partial^2_{\hat{r}}\hat{\chi}+\frac{d-1}{\hat{r}}\partial_{\hat{r}}\hat{\chi}+\partial_{\hat z}^2\hat \chi &=(1+\hat{\varphi})\hat{\chi} \ ,\\
 	\partial^2_{\hat{r}}\hat{\varphi}+\frac{d-1}{\hat{r}}\partial_{\hat{r}}\hat{\varphi}+\partial_{\hat z}^2\hat \varphi &=\hat{\chi}^2\ .
	\end{split}
	\end{equation}
Note that in these rescaled variables (\ref{resHP}), the periodicity of the $z$ circle changes to $\hat z\sim \hat z+m_\infty L$.
\subsection{Uniform solutions}
For solutions that are uniform along the $z$ circle, meaning the fields $\chi$ and $\varphi$ are independent of $z$, the e.o.m. (\ref{eomHPres}) reduce to the following ordinary differential equations.
\begin{equation}
\label{eomHPresu}
	\begin{split}
 	\hat{\chi}''(\hat r)+\frac{d-1}{\hat{r}}\hat{\chi}'(\hat r) &=(1+\hat{\varphi})\hat{\chi} \ ,\\
 	\hat{\varphi}''(\hat r)+\frac{d-1}{\hat{r}}\hat{\varphi}'(\hat r) &=\hat{\chi}^2\ .
	\end{split}
	\end{equation}
At the action level, for uniform fields, the integration over $z$ in (\ref{Sphichiz}) recovers the $d$-dimensional effective action (\ref{Sphichid}).

The differential equations (\ref{eomHPresu}) can be solved numerically using the shooting method. This involves imposing the initial conditions $\hat\chi'(0)=\hat\varphi'(0)=0$ and fine-tuning the values of $\hat\chi(0)$ and $\hat\varphi(0)$ to obtain solutions where both $\hat\chi$ and $\hat\varphi$ asymptotically approach zero (see, for example,~\cite{Balthazar:2022szl}).

Substituting the HP solutions $\chi(r)$ and $\varphi(r)$ into the action yields the free energy of the canonical ensemble. Specifically, from (\ref{Sphichid}), we have
\begin{equation}
\label{FId}
\begin{split}
    F=\frac{I_d}{\beta}=& \frac{\omega_{d-1}}{16\pi G_N^{(d+1)}}\int dr\ r^{d-1}\left[(\nabla\varphi)^2+(\nabla \chi)^2+\left(m_\infty^2+\frac{\kappa}{\alpha'}\varphi\right)\chi^2\right]\\
    =& \frac{\omega_{d-1}}{16\pi G_N^{(d+1)}}\int dr\ r^{d-1}\left[-\varphi\nabla^2\varphi-\chi\nabla^2 \chi+\left(m_\infty^2+\frac{\kappa}{\alpha'}\varphi\right)\chi^2\right]\\
    =& \frac{\omega_{d-1}}{32\pi G_N^{(d+1)}}\frac{\kappa}{\alpha'}\int dr\ r^{d-1}\left(-\varphi\chi^2\right)\ .
\end{split}
\end{equation}
Note that in the second line we have integrated by parts and in the third line we have used the equations of motion. Using the rescaled variables and fields (\ref{resHP}), the free energy can be explicitly expressed in terms of $m_\infty$ as
\begin{equation}
\label{Fm}
    F=\frac{\omega_{d-1}m_\infty^{6-d}}{16\pi G_N^{(d+1)}}\left(\frac{\alpha'}{\kappa}\right)^2\int d\hat r\ \hat r^{d-1}\left(-\hat\varphi\hat\chi^2\right)\ .
\end{equation}

With the HP solution, we can also calculate the thermodynamic quantities of the microcanonical ensemble. First, the ADM mass can be determined from the asymptotic form of $\varphi(r)$, i.e.,
\begin{equation}
\label{asym}
   \varphi(r)\sim -\frac{C_\varphi}{r^{d-2}}\ .
\end{equation}
By integrating the second equation in (\ref{eomHP}), the fall-off coefficient $C_\varphi$ is found to be
\begin{equation}
\label{intchi}
    C_\varphi=\frac{\kappa}{2(d-2)\alpha'}\int_0^\infty dr\ r^{d-1}\chi^2(r)\ .
\end{equation}
Thus, the mass of the HP solution can be expressed as
\begin{equation}
\label{MC}
    M=\frac{(d-2)\omega_{d-1}}{8\pi G_N^{(d+1)}}C_\varphi=\frac{\omega_{d-1}}{16\pi G_N^{(d+1)}}\frac{\kappa}{\alpha'}\int_0^\infty dr\ r^{d-1}\chi^2(r)\ .
\end{equation}

Using the rescaling in (\ref{resHP}), we can further extract the dependence of the mass on $m_\infty$, which is given by
\begin{equation}
\label{Mm}
    M=\frac{\omega_{d-1}m_\infty^{4-d}}{8\pi G_N^{(d+1)}}\frac{\alpha'}{\kappa}\int_0^\infty d\hat r\ \hat r^{d-1}\hat\chi^2(\hat r)\ .
\end{equation}
This expression reveals that for $d<4$, $M$ increases with $m_\infty$, while for $d>4$, $M$ decreases with $m_\infty$. In the special case of $d=4$, determining the dependence of $M$ on $m_\infty$ requires including subleading terms in the effective action. We will come back to this issue in section \ref{qua}.

For a given mass $M$, the entropy can be calculated using the formula
\begin{equation}
\label{SMF}
    S=\beta M-\beta F\ .
\end{equation}
Expanding this expression to leading orders in $\beta-\beta_H$ yields
\begin{equation}
\label{SHP}
\begin{split}
    S=&\beta_HM+(\beta-\beta_H)M-\beta_HF+O\left((\beta-\beta_H)^2\right)\\
    =&\beta_HM+\frac{2(\pi\alpha')^2}{\beta_H}m_\infty^2M-\beta_HF+O\left((\beta-\beta_H)^2\right)\ ,
\end{split}
\end{equation}
where $m_\infty$ is related to $M$ as in (\ref{Mm}).
\subsection{Non-uniform solutions}\label{secnu}
More generally, for solutions that are non-uniform along the $z$ circle, we need to solve the full partial differential equations (\ref{eomHPres}). As pointed out by~\cite{Emparan:2024mbp}, the relaxation method is more practical for numerically solving these equations than the shooting method. Besides, it is useful to compactify the $\hat r$ coordinate as
\begin{equation}
    \hat u=\left(\frac{\hat r}{1+\hat r}\right)^2\ ,
\end{equation}
which maps the domain $\hat r\in [0,\infty)$ to $\hat u\in [0,1]$~\cite{Emparan:2024mbp}. This domain is then discretized using a Lobatto-Chebyshev grid (see, e.g., Appendix A of \cite{Dias:2015nua} for technical details). For the $\hat z$ coordinate, reflection symmetry allows us to focus on the half-period $\hat z\in [0,m_\infty L/2]$, which is similarly discretized using a Lobatto-Chebyshev grid. Additionally, we impose the boundary conditions as 
\begin{equation}
\label{bc}
    \partial_{\hat{z}} \hat{\chi}(\hat u,\hat{z})|_{\hat z=0,\frac{m_\infty L}{2}}=\partial_{\hat{z}}\hat{\varphi} (\hat u,\hat z)|_{\hat z=0,\frac{m_\infty L}{2}}=\hat\chi(\hat u,\hat z)|_{\hat u=1}=\hat\varphi(\hat u,\hat z)|_{\hat u=1}=0\ .
\end{equation}
\begin{figure}
	\centering
	\subfigure[]{
	\begin{minipage}[t]{0.3\linewidth}
	\centering
	\includegraphics[width=1.8in]{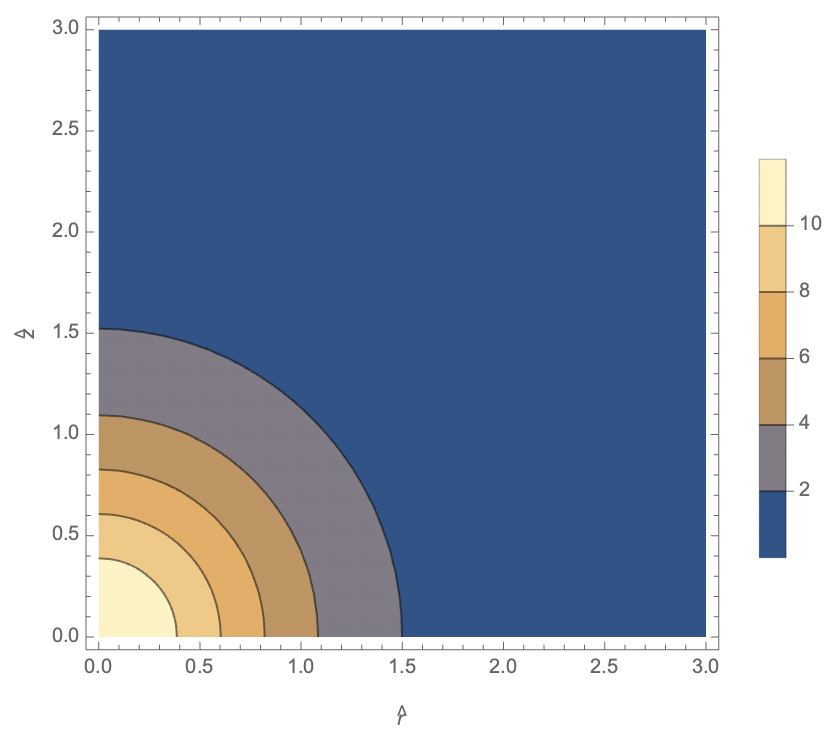}\label{NUHPzm3}
	\end{minipage}}
	\subfigure[]{
	\begin{minipage}[t]{0.3\linewidth}
	\centering
	\includegraphics[width=1.8in]{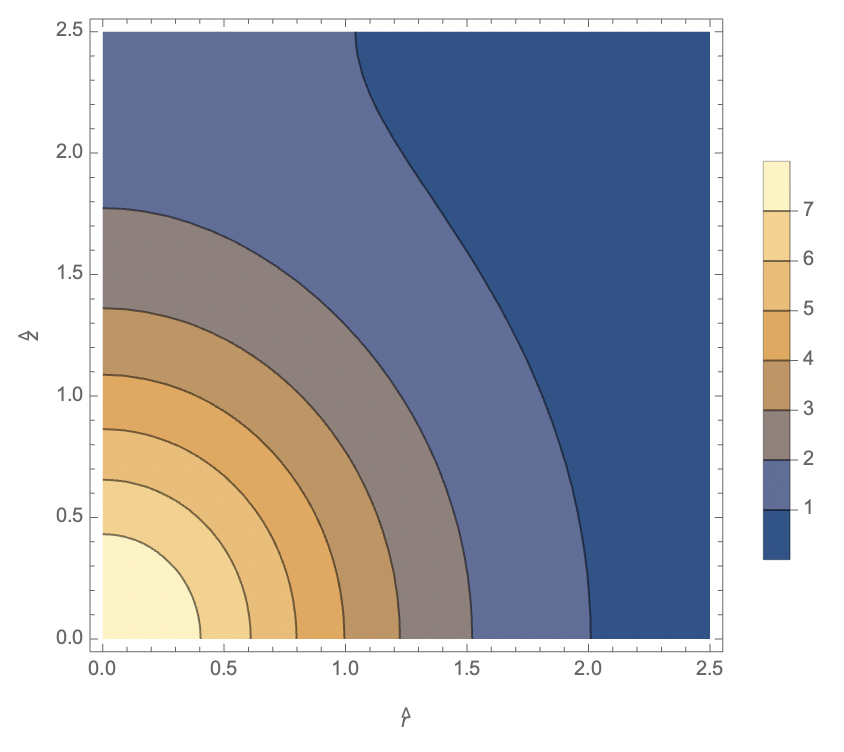}\label{NUHPzm25}
	\end{minipage}}
    \subfigure[]{
	\begin{minipage}[t]{0.3\linewidth}
	\centering
	\includegraphics[width=1.8in]{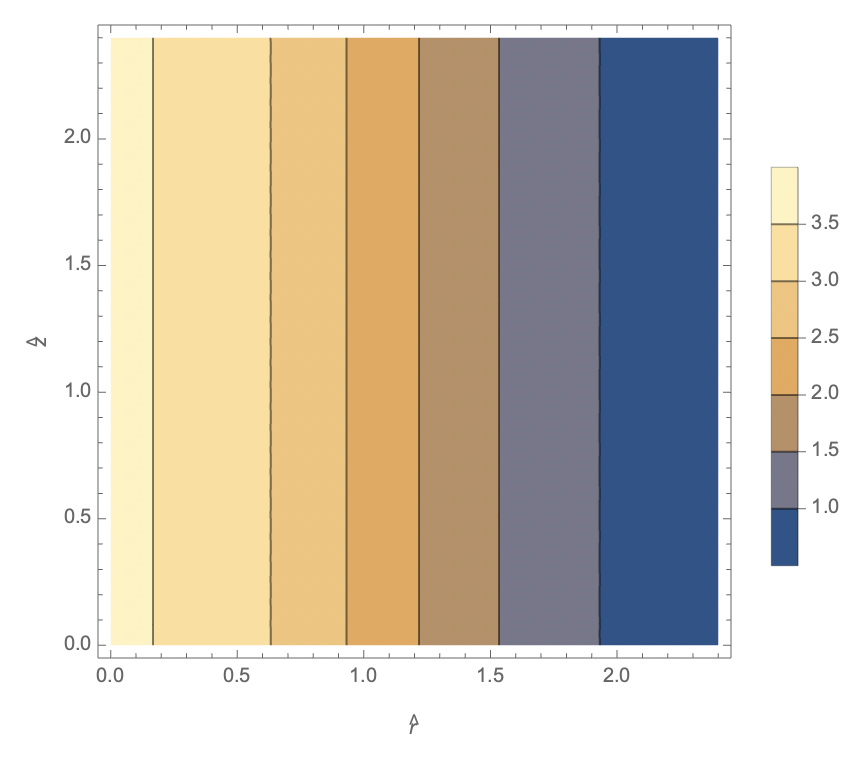}\label{NUHPzm24}
	\end{minipage}}
	\centering
\caption{\label{NUHPd4}Contour plots of $\hat\chi(\hat r,\hat z)$ for $d=4$ and (a) $m_\infty L=6$, (b) $m_\infty L=5$, (c) $m_\infty L=4.8$.}
\end{figure}

To apply the relaxation method, an initial seed solution is needed. The rescaling in (\ref{resHP}) implies that the size of the solution is characterized by $1/m_\infty$. Thus, when $m_\infty$ is large compared to $1/L$, non-uniform solutions are localized in the $z$ circle. These solutions can be approximated by the spherically symmetric solutions in the noncompact space $\mathbb{R}^{d+1}$, i.e., the solutions of (\ref{eomHPresu}) with $d$ replaced by $d+1$. The latter can be obtained via the shooting method. With $\hat r$ replaced by $\sqrt{\hat r^2+\hat z^2}$, they then serve as seeds for the relaxation method. Furthermore, once a solution of (\ref{eomHPres}) is found, it can be iteratively used as a new seed to generate more solutions for progressively smaller values of $m_\infty L$.

In figure \ref{NUHPd4}, we present the numerical solutions of $\hat\chi(\hat r,\hat z)$, for $d=4$ as an example. The plots show that when $m_\infty L$ is sufficiently large, the HP solution is indeed localized in $\mathbb{S}_z^1$. As $m_\infty L$ decreases, the solution becomes increasingly uniform. At a critical value of $m_\infty$ between $4.8/L$ and $5/L$, the non-uniform solution transitions into a uniform solution. 

Also note that the amplitudes of the solutions are of order 1 (e.g., figure \ref{NUHPd4}). Consequently, from (\ref{resHP}), the amplitudes of the unhatted fields $\chi$ and $\varphi$ scale as $\alpha'm_\infty^2$. The EFT is valid only when the fields are small, which requires $\alpha'm_\infty^2\ll 1$. Therefore, for the solutions to remain valid near the critical point, which corresponds to $m_\infty=O(L^{-1})$, the condition $L\gg l_s$ must be satisfied—precisely the regime of interest in this paper.

With the solutions for $\hat\chi(\hat r,\hat z)$ and $\hat\varphi(\hat r,\hat z)$, we can compute thermodynamic quantities such as the free energy, mass, and entropy. First, similar to (\ref{Fm}), the free energy of the canonical ensemble is given by
\begin{equation}
\label{F}
    F=\frac{\omega_{d-1}L^{d-5}}{8\pi G_N^{(d+2)}}\left(\frac{\alpha'}{\kappa}\right)^2f\ ,
\end{equation}
where
\begin{equation}
\label{f}
    f\equiv (m_\infty L)^{5-d}\int_0^{\frac{m_\infty L}{2}}d\hat z\int_0^\infty d\hat r\ \hat r^{d-1}\left(-\hat\varphi\hat\chi^2\right)\ .
\end{equation}

For the microcanonical ensemble, the mass of the HP solution can be generalized from (\ref{Mm}) to
\begin{equation}
\label{M}
    M=\frac{\omega_{d-1}}{4\pi  G_N^{(d+2)}}\frac{\alpha'}{\kappa}L^{d-3}m\ ,
\end{equation}
where $m$ is defined as
\begin{equation}
\label{m}
    m\equiv (m_\infty L)^{3-d}\int_0^{\frac{m_\infty L}{2}}d\hat z\int_0^\infty d\hat r\ \hat r^{d-1}\hat \chi^2(\hat r,\hat z)\ .
\end{equation}

With the mass $M$ (\ref{M}) and the free energy $F$ (\ref{F}), we can further calculate the entropy. More specifically, using (\ref{SHP}), we obtain
\begin{equation}
\label{SMFz}
\begin{split}
    S=\beta_HM+\frac{\omega_{d-1}\pi^3\alpha'^4}{\beta_H^3G_N^{(d+2)}}L^{d-5}s\ ,
\end{split}
\end{equation}
where
\begin{equation}
\label{s}
    s\equiv (m_\infty L)^2m-\frac{f}{2}\ . 
\end{equation}
Here, the first term in (\ref{SMFz}) reflects the Hagedorn behavior, while the second term represents the leading-order correction to it.

\subsection{$d=2$}\label{secd2}
We are motivated to investigate the HP solutions for $d=2$ due to the interesting behavior of black holes in this context. In particular, it has been found that as the size increases and the horizon becomes uniform, a black hole would expand infinitely, swallowing the whole universe~\cite{Bogojevic:1990hv,Frolov:2003kd}. Thus, it is intriguing to study what happens when we start with a localized HP solution at a sufficiently large $m_\infty$ and gradually decrease $m_\infty$, thereby increasing the size of the solution.\footnote{I am grateful to Roberto Emparan, Mikel Sanchez-Garitaonandia and Marija Toma\v{s}evi\'{c} for bringing this to my attention and for providing valuable insights on the topic.}

However, a subtlety arises in $\mathbb{R}^2\times \mathbb{S}_z^1$. For any solution of $\hat\chi$ that vanishes asymptotically, the second line of the e.o.m. (\ref{eomHPres}) implies that $\hat\varphi$ diverges logarithmically at large $\hat r$. In other words, the circumference of the Euclidean time circle cannot converge to a constant, meaning that the canonical and microcanonical ensembles may not be well-defined. Note that this situation also occurs in the black hole case. 

Nevertheless, we expect that bounded solutions for $\hat\chi$ can still be found. To investigate this, as described in the last subsection, we begin with a localized HP solution as a seed for the relaxation method and look for non-uniform solutions of (\ref{eomHPres}) by progressively decreasing $m_\infty L$. Note that since the Euclidean time circle does not asymptotically approach a constant, $m_\infty$ no longer carries a direct physical interpretation, such as representing the asymptotic mass of the field $\chi$. However, it remains a convenient parameter for characterizing the solutions.

The logarithmic divergence of $\hat \varphi$ at large $\hat r$ prevents the field from satisfying the boundary conditions given in (\ref{bc}), complicating the numerical analysis. To address this, we redefine the field as
\begin{equation}
    \hat \varphi=(1+\hat r^2)\hat \phi\ ,
\end{equation}
and accordingly modify the boundary conditions to
\begin{equation}
\label{bc2}
    \partial_{\hat{z}} \hat{\chi}(\hat u,\hat{z})|_{\hat z=0,\frac{m_\infty L}{2}}=\partial_{\hat{z}}\hat{\phi} (\hat u,\hat z)|_{\hat z=0,\frac{m_\infty L}{2}}=\hat\chi(\hat u,\hat z)|_{\hat u=1}=\hat\phi(\hat u,\hat z)|_{\hat u=1}=0\ .
\end{equation}
\begin{figure}
	\centering
	\subfigure[]{
	\begin{minipage}[t]{0.45\linewidth}
	\centering
	\includegraphics[width=2.5in]{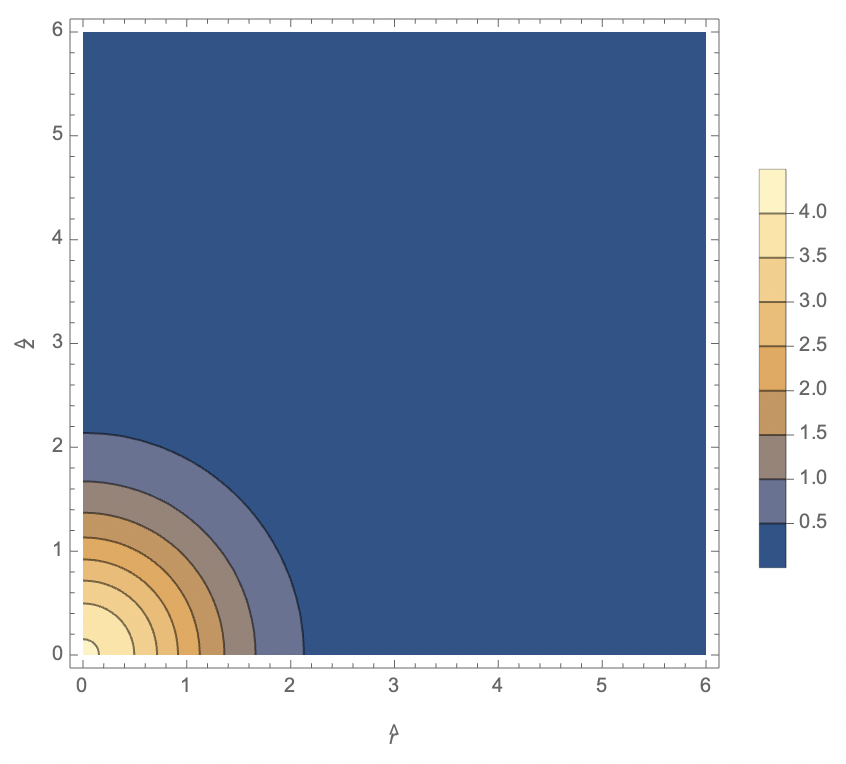}\label{chid2zm6}
	\end{minipage}}
        \subfigure[]{
	\begin{minipage}[t]{0.45\linewidth}
	\centering
	\includegraphics[width=2.5in]{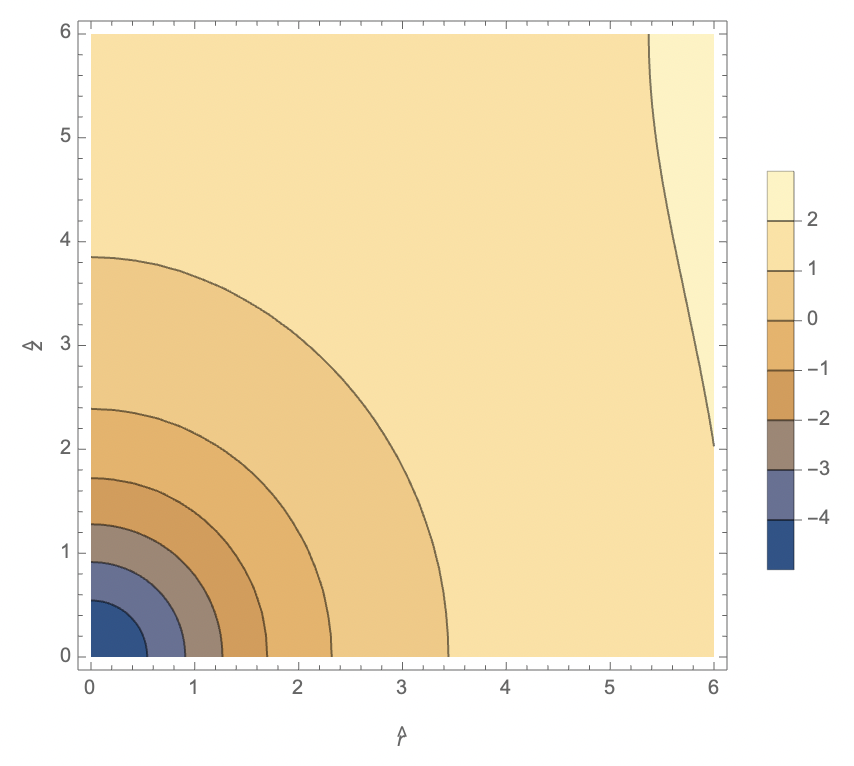}\label{phid2zm6}
	\end{minipage}}
	\centering
\caption{\label{chiphid2zm6}Non-uniform solutions of (a) $\hat\chi$ and (b) $\hat\varphi$ for $d=2$, with $m_\infty L=12$.}
\end{figure}
\begin{figure}
	\centering
\includegraphics[scale=0.6]{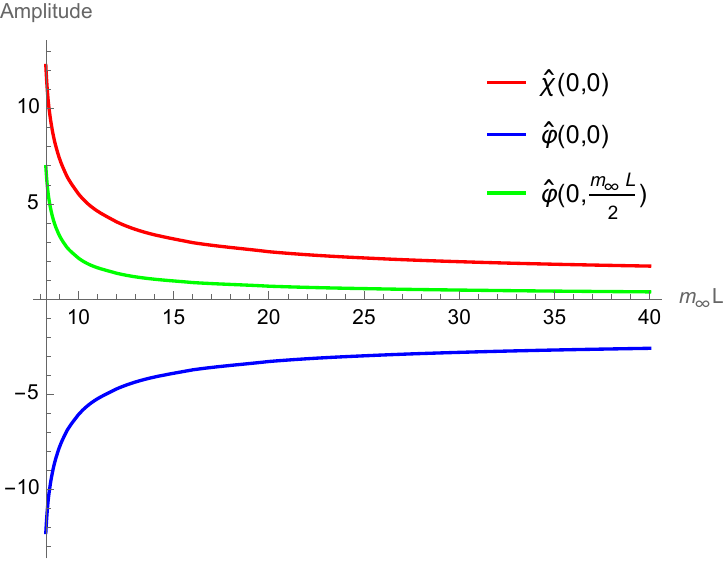}
\caption{\label{ampzmd2}$\hat \chi (\hat r=0,\hat z=0)$, $\hat \varphi (\hat r=0,\hat z=0)$ and $\hat \varphi (\hat r=0,\hat z=\frac{m_\infty L}{2})$ as functions of $m_\infty L$.}
\end{figure}

Following the above procedures, we obtain solutions where $\chi$ is bounded. As shown in figure \ref{chiphid2zm6}, with $m_\infty L=12$ as an example, $\hat\varphi$ does diverge to positive infinity at large $\hat{r}$, but $\hat\chi$ goes to zero asymptotically.

Besides, it turns out that as $m_\infty$ decreases, the non-uniform solution does not transition into a uniform solution, similar to the behavior observed in the black hole case. In fact, as shown in figure \ref{ampzmd2}, the field amplitudes diverge as $m_\infty$ approaches a constant. Especially, $\hat\varphi (0,0)$ diverges to negative infinity, while $\hat\varphi (0,m_\infty L/2)$ diverges to positive infinity, which indicates that the non-uniform solution never becomes uniform. 

\begin{figure}
	\centering
	\subfigure[]{
	\begin{minipage}[t]{0.45\linewidth}
	\centering
	\includegraphics[width=2.5in]{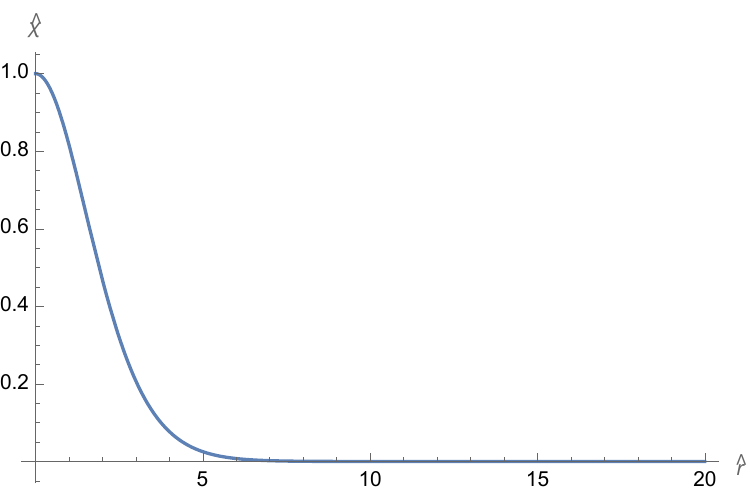}\label{chid2}
	\end{minipage}}
        \subfigure[]{
	\begin{minipage}[t]{0.45\linewidth}
	\centering
	\includegraphics[width=2.5in]{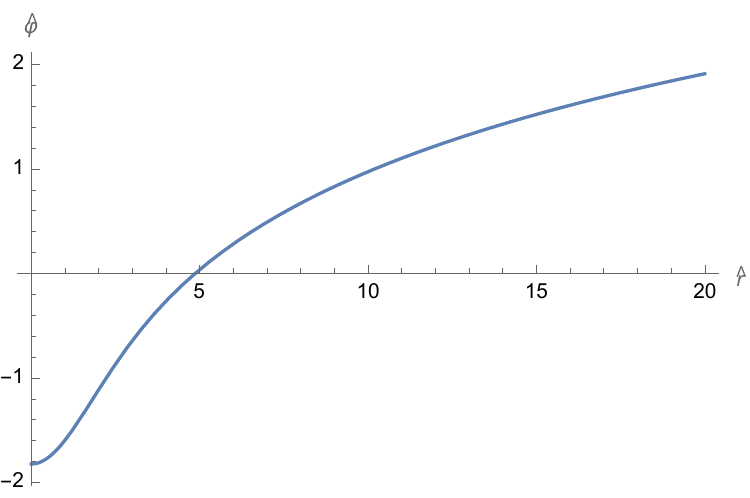}\label{phid2}
	\end{minipage}}
	\centering
\caption{\label{chiphid2}Uniform HP solutions $\hat\chi$ and $\hat\varphi$ for $d=2$ with $\hat\chi(0)=1$.}
\end{figure}

However, this does not imply that uniform solutions are absent. By numerically solving the e.o.m. (\ref{eomHPresu}), we do find uniform solutions where $\hat \chi$ vanishes asymptotically. For instance, figure \ref{chiphid2} shows a solution with $\hat\chi(0)=1$.

Unlike the HP solutions for $d>2$, the uniform solution here is not unique. This is because the boundary condition $\hat\varphi=0$ at infinity is no longer applicable for $d=2$. As a result, any scaling transformation of the solution shown in figure \ref{chiphid2}, with
\begin{equation}
\label{scsy}
    \hat{\chi}\to \lambda^2\hat{\chi}\ ,\quad \hat{\varphi}\to \lambda^2(\hat{\varphi}+1)-1\ ,\quad\hat{r}\to \frac{\hat{r}}{\lambda}\ ,
\end{equation}
yields another solution of (\ref{eomHPresu}).

This scaling symmetry also explains why the non-uniform solution does not transition into the uniform solution as $m_\infty$ varies. Suppose such a transition does occur at some critical value $m_{\infty,c}L$. Then, the transformation (\ref{scsy}), combined with $\hat z\to \hat z/\lambda$, under which the e.o.m. (\ref{eomHPres}) is invariant, would map the critical value to $m_{\infty,c}L/\lambda$ (recall that $\hat z\sim \hat z+m_\infty L$). Thus, the critical point becomes ambiguous, making the transition scenario problematic.

\begin{figure}
	\centering
        \subfigure[]{
	\begin{minipage}[t]{0.3\linewidth}
	\centering
	\includegraphics[width=1.8in]{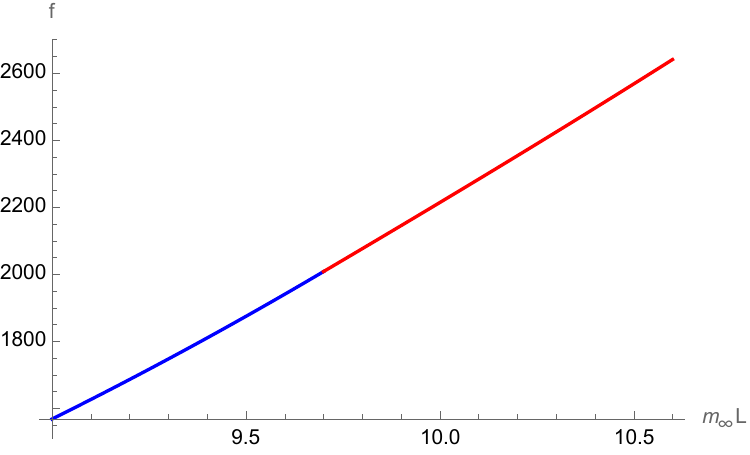}\label{fzd27}
	\end{minipage}}
	\subfigure[]{
	\begin{minipage}[t]{0.3\linewidth}
	\centering
	\includegraphics[width=1.8in]{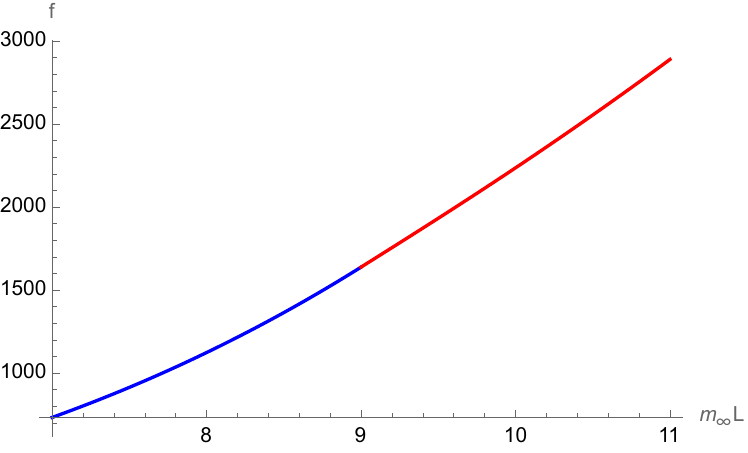}\label{fzd28}
	\end{minipage}}
        \subfigure[]{
	\begin{minipage}[t]{0.3\linewidth}
	\centering
	\includegraphics[width=1.8in]{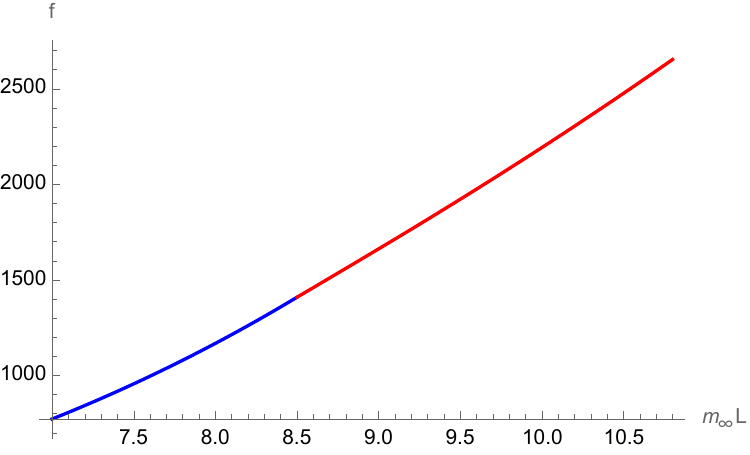}\label{fzd29}
	\end{minipage}}
	\centering
\caption{\label{fz2d3}Phase diagrams of the canonical ensemble for (a) $d=2.7$, (b) $d=2.8$ and (c) $d=2.9$. The red curves represent non-uniform solutions, and the blue curves represent uniform solutions.}
\end{figure}
\begin{figure}
	\centering
        \subfigure[]{
	\begin{minipage}[t]{0.3\linewidth}
	\centering
	\includegraphics[width=1.8in]{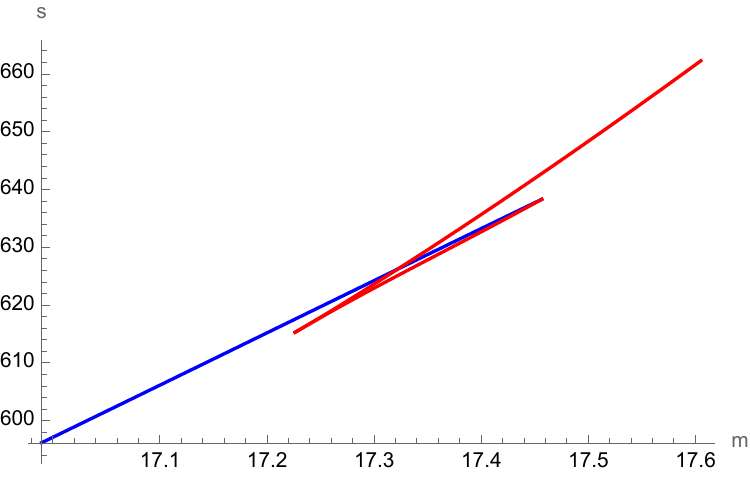}\label{smd27}
	\end{minipage}}
	\subfigure[]{
	\begin{minipage}[t]{0.3\linewidth}
	\centering
	\includegraphics[width=1.8in]{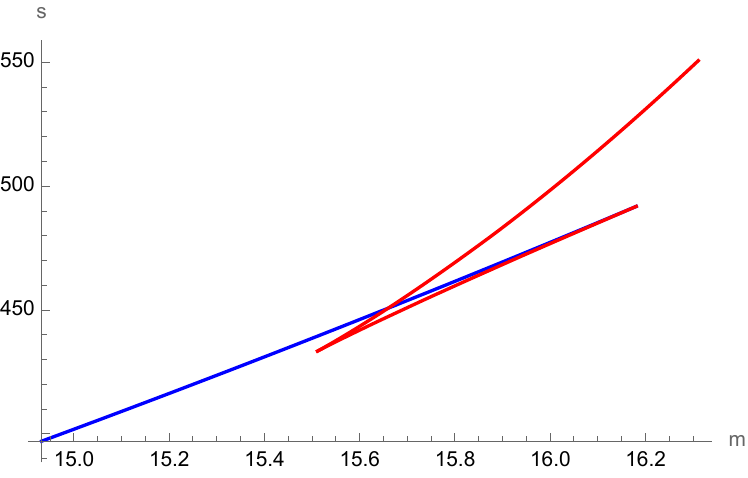}\label{smd28}
	\end{minipage}}
        \subfigure[]{
	\begin{minipage}[t]{0.3\linewidth}
	\centering
	\includegraphics[width=1.8in]{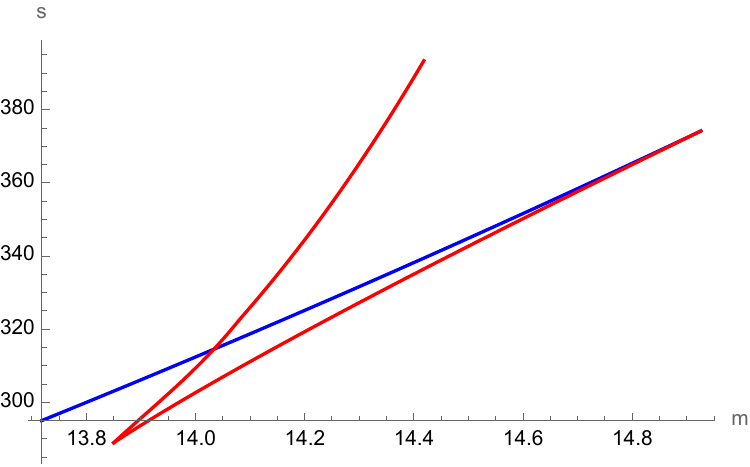}\label{smd29}
	\end{minipage}}
	\centering
\caption{\label{sm2d3}Phase diagrams of the microcanonical ensemble for (a) $d=2.7$, (b) $d=2.8$ and (c) $d=2.9$.}
\end{figure}
Finally, note that the case of $d=2$ differs qualitatively from cases where $d$ lies between 2 and 3. In particular, for $2<d<3$, $\hat \varphi$ vanishes asymptotically, allowing the canonical and microcanonical ensembles to be well-defined. This also enables us to impose the boundary conditions (\ref{bc}) rather than (\ref{bc2}). 

Taking $d=2.7$, 2.8 and 2.9 as examples, we begin with localized solutions and progressively decrease $m_\infty L$ to generate additional solutions with increasingly uniformity. The rescaled free energy $f$ (defined in (\ref{f})) as a function of $m_\infty L$ is plotted in figure \ref{fz2d3}. Unlike the case of $d=2$, as $m_\infty$ decreases, the non-uniform HP solution transitions into a uniform solution at a critical point. Since the free energy change is continuous at the critical point, the phase transition can be classified as second-order.

For the microcanonical ensemble, the phase diagrams are shown in figure \ref{sm2d3}. For each $d$, there are two branches of non-uniform solutions. Along with the uniform branch, they form a swallowtail-type phase diagram. This indicates that the phase transition between the uniform and non-uniform solutions is first-order, consistent with the perturbative analysis in~\cite{Chu:2024ggi}.
\subsection{$d=3$ and $d=4$}\label{secd34}
\begin{figure}
	\centering
	\subfigure[]{
	\begin{minipage}[t]{0.45\linewidth}
	\centering
	\includegraphics[width=2.5in]{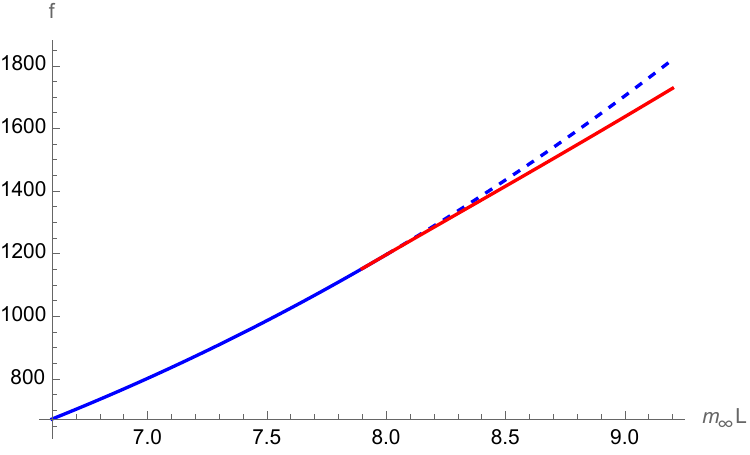}\label{fzd3}
	\end{minipage}}
	\subfigure[]{
	\begin{minipage}[t]{0.45\linewidth}
	\centering
	\includegraphics[width=2.5in]{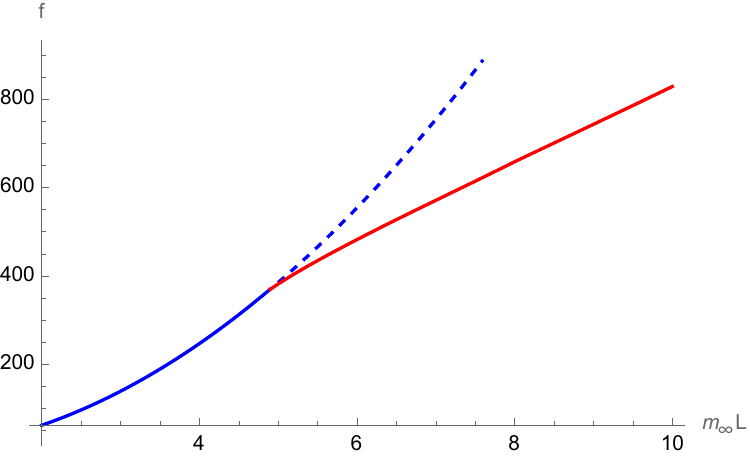}\label{fzd4}
	\end{minipage}}
	\centering
\caption{\label{cpdd34}Phase diagrams of the HP solutions in the canonical ensemble for (a) $d=3$ and (b) $d=4$. The red curves denote the non-uniform solutions, and the blue curves represent the uniform solutions.}
\end{figure}
The phase diagrams of the canonical ensemble for $d=3$ and $d=4$ are presented in figure \ref{cpdd34}, which recover the results of~\cite{Emparan:2024mbp}. Similar to the case of $2<d<3$, the non-uniform solution undergoes a second-order phase transition into the uniform solution at the critical point ($m_\infty \approx 7.9/L$ for $d=3$ and $m_\infty \approx 4.8/L$ for $d=4$).

We also include in figure \ref{cpdd34} the uniform solutions above the critical value of $m_\infty$, represented by the dashed blue curve. It can be seen that the free energies of these uniform solutions are higher than those of the non-uniform solutions, indicating that they are unstable towards non-uniformity. Notably, they lie at $m_\infty$ larger than the critical value, where the size of the solutions in the direction of $r$, characterized by $m_\infty^{-1}$, is small compared to $L$.
\begin{figure}
	\centering
	\subfigure[]{
	\begin{minipage}[t]{0.45\linewidth}
	\centering
	\includegraphics[width=2.5in]{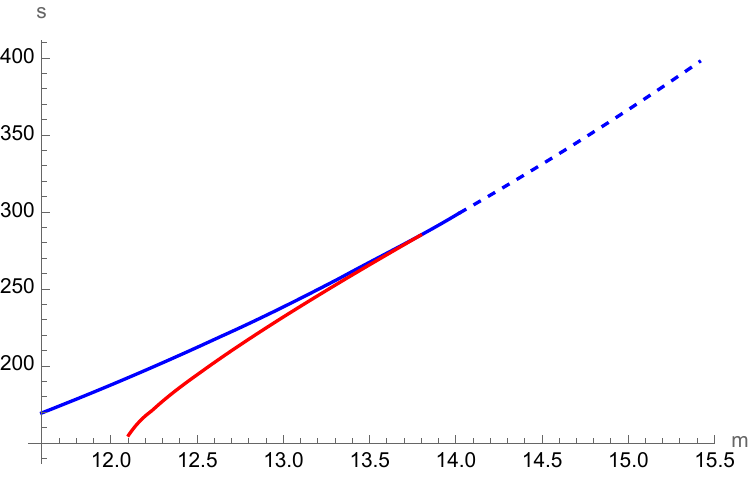}\label{smd3}
	\end{minipage}}
	\subfigure[]{
	\begin{minipage}[t]{0.45\linewidth}
	\centering
	\includegraphics[width=2.5in]{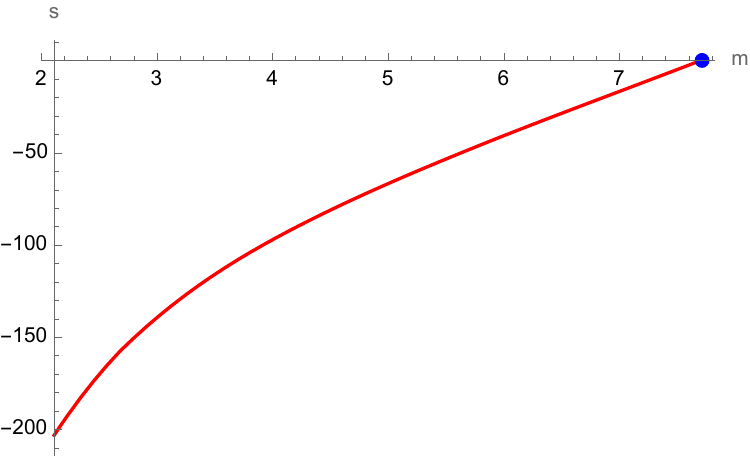}\label{smd4}
	\end{minipage}}
	\centering
\caption{\label{mcpdd34}Phase diagrams of the HP solutions in the microcanonical ensemble for (a) $d=3$ and (b) $d=4$. The red curves correspond to non-uniform solutions, while the blue curves and dot represent uniform solutions.}
\end{figure}

In the microcanonical ensemble, the phase diagrams for $d=3$ and $d=4$ differ qualitatively from those for $2<d<3$ (e.g., figure \ref{sm2d3}). First, for $d=3$, as shown in figure \ref{smd3}, the non-uniform solution always has a lower entropy than the uniform solution (see also \cite{Emparan:2024mbp}). Thus, the uniform solution can never transition into a non-uniform HP solution. 

However, this raises the question of what state the uniform HP solution transitions into at the critical mass. Notably, as $m_\infty$ increases, the non-uniform HP solution decreases in mass and becomes increasingly localized. When $m_\infty$ approaches $O(l_s^{-1})$, the HP EFT breaks down, and this is supposed to be the regime where the localized HP solution transitions into a localized black hole. It is then reasonable that the black hole branch complements figure \ref{smd3} to form a swallowtail-type phase diagram, similar to those in figure \ref{sm2d3}. In other words, the uniform HP solution undergoes a first-order phase transition into a localized black hole at the critical mass, as proposed in~\cite{Chu:2024ggi}.

For $d=4$, the mass of the uniform HP solution is fixed, as discussed before. Thus, the uniform solution is denoted by a blue dot in figure \ref{smd4}. To determine how the mass varies with $m_\infty$, it is necessary to include the quartic terms in the effective action, which we now analyze in the next section.
\section{Beyond HP EFT}\label{secHP4}
As discussed in~\cite{Chen:2021dsw}, the HP solution in $\mathbb{R}^d$ ceases to exist for $d\ge 6$. Furthermore, ~\cite{Balthazar:2022hno,Chu:2024ggi} show that when $d$ is close to 6 as a continuous parameter, the effective action (\ref{Sphichid}) breaks down because the quartic terms become significant. However, the higher-order terms, such as $|\chi|^6$, are still suppressed as long as $\chi,\varphi\ll 1$. In this section, we extend the original HP analysis by incorporating these quartic terms, which are derived from string scattering amplitudes in~\cite{Brustein:2021ifl} (for bosonic and type II strings), leading to the following corrected action.
\begin{equation}
\label{Id1}
	\begin{split}
 	I_{d+1}=\frac{\beta}{16\pi G_N^{(d+2)}}\int_{-\frac{L}{2}}^{\frac{L}{2}}dz\int d^dx\bigg[&\left(\partial_z\varphi\right)^2+(\nabla_x\varphi)^2+\left(\partial_z\chi\right)^2+|\nabla_x \chi|^2\\
    &+\left(m_\infty^2+\frac{\kappa}{\alpha'}\varphi+\frac{\kappa}{\alpha'}\varphi^2\right)|\chi|^2+\frac{\kappa}{4\alpha'}|\chi|^4\bigg]\ .
	\end{split}
	\end{equation}
    
We will use this quartic-corrected action to study string stars for $d=5$ and $d=6$. Before proceeding, however, let us first explore how the quartic terms determine the dependence of $M$ on $m_\infty$ for the uniform solution at $d=4$.
\subsection{More on $d=3$ and $d=4$}\label{qua}
For the uniform solution, the e.o.m. derived from the quartic-corrected action (\ref{Id1}) are given by
\begin{equation}
 \label{eomHP4}
	\begin{split}
 	\chi''(r)+\frac{d-1}{r}\chi'(r)&=\left(m_{\infty}^2+\frac{\kappa}{\alpha'}\varphi(r)+\frac{\kappa}{\alpha'}\varphi^2(r)+\frac{\kappa}{2
    \alpha'}\chi^2(r)\right)\chi(r) \ ,\\
 	\varphi''(r)+\frac{d-1}{r}\varphi'(r) &=\frac{\kappa}{2\alpha'}\left(1+2\varphi(r)\right)\chi^2(r)\ .
	\end{split}
	\end{equation}

To analyze corrections to the HP solution, we expand the fields perturbatively as
\begin{equation}
\label{exp}
	\begin{split}
 	\chi &=\frac{\sqrt{2}\alpha' }{\kappa}m^2_{\infty}\left(\hat{\chi}+\frac{\alpha' }{\kappa}m^2_{\infty}\hat{\hat{\chi}}+O((\alpha'm_\infty^2)^2)\right)\ ,\\
 	\varphi &=\frac{\alpha'}{\kappa }m^2_{\infty}\left(\hat{\varphi}+\frac{\alpha' }{\kappa}m^2_{\infty}\hat{\hat{\varphi}}+O((\alpha'm_\infty^2)^2)\right)\ .
	\end{split}
	\end{equation}
Substituting this expansion into (\ref{eomHP4}) and using the rescaled variable $\hat r=m_\infty r$, we derive the leading-order ($O(m_\infty^2)$) and subleading-order ($O(m_\infty^4)$) differential equations. Specifically, the $O(m_\infty^2)$ equations govern $\hat \chi$ and $\hat \varphi$, and are identical to those in (\ref{eomHPresu}). The $O(m_\infty^4)$ equations, which govern $\hat{\hat\chi}$ and $\hat{\hat \varphi}$, are given by
\begin{equation}
\label{eomHPresres}
	\begin{split}
 	\hat{\hat \chi}''(\hat r)+\frac{d-1}{\hat{r}}\hat{\hat \chi}'(\hat r) &=(1+\hat{\varphi}(\hat r)\hat{\hat\chi}(\hat r)+\hat\chi (\hat r)\hat{\hat \varphi}(\hat r)+\hat\chi^3(\hat r)+\hat\varphi^2(\hat r)\hat\chi (\hat r)\ ,\\
 	\hat{\hat \varphi}''(\hat r)+\frac{d-1}{\hat{r}}\hat{\hat \varphi}'(\hat r)&=2\hat\chi(\hat r)\hat{\hat\chi}(\hat r)+2\hat\chi^2(\hat r)\hat\varphi (\hat r)\ .
	\end{split}
	\end{equation}
\begin{figure}
	\centering
	\subfigure[]{
	\begin{minipage}[t]{0.45\linewidth}
	\centering
	\includegraphics[width=2.5in]{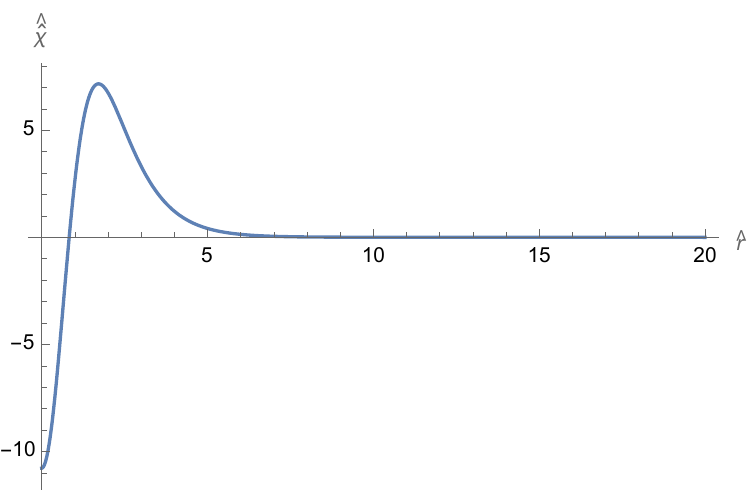}\label{hathatchi}
	\end{minipage}}
	\subfigure[]{
	\begin{minipage}[t]{0.45\linewidth}
	\centering
	\includegraphics[width=2.5in]{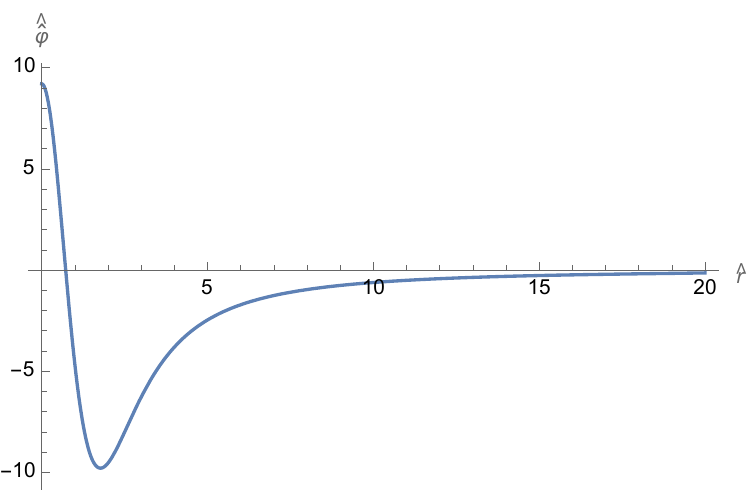}\label{hathatphi}
	\end{minipage}}
	\centering
\caption{\label{hathatchiphi}Profiles of the leading-order perturbations $\hat{\hat\chi}$ and $-\hat{\hat\varphi}$ to the Horowitz-Polchinski solution at $d=4$.}
\end{figure}

As before, we can first use the shooting method to numerically solve (\ref{eomHPresu}) for $\hat \chi$ and $\hat\varphi$. Once the solutions are obtained, they can be plugged in (\ref{eomHPresres}), and the shooting method can be applied again to solve for the fields $\hat{\hat\chi}$ and $\hat{\hat \varphi}$. For $d=4$, the solutions are plotted in figure \ref{hathatchiphi}. Notably, while $\hat{\hat\varphi}$ is positive at small $\hat r$, it eventually becomes negative as $\hat r$ increases. In other words, the fall-off coefficient $\hat{\hat C}_\varphi$ in the asymptotic behavior $\hat{\hat\varphi}$, given by
\begin{equation}
\label{hathatCphi}
    \hat{\hat\varphi}\sim -\frac{\hat{\hat C}_\varphi}{\hat r^{d-2}}\ ,
\end{equation}
is positive. 

We similarly define the fall-off coefficient of $\hat\varphi$ as $\hat{C}_\varphi$, such that $\hat\varphi\sim -\hat{C}_\varphi/\hat r^{d-2}$. Then, combining (\ref{exp}), (\ref{hathatCphi}) and $r=\hat r/m_\infty$, the fall-off coefficient of $\varphi$ in (\ref{asym}) can be written as
\begin{equation}
\label{Cm}
    C_\varphi=\frac{\alpha'm_\infty^{4-d}}{\kappa }\left(\hat{C}_\varphi+\frac{\alpha' }{\kappa}m^2_{\infty}\hat{\hat{C}}_\varphi+O((\alpha'm_\infty^2)^2)\right).
\end{equation}
Since $\hat{\hat{C}}_\varphi>0$ for $d=4$, it follows that the mass of the uniform solution increases with $m_\infty$.

With this new result, we now provide more comments on the phase diagrams of the microcanonical ensembles for $d=3$ and $d=4$. Combining figure \ref{fzd3} with figure \ref{smd3}, we can deduce that for $d=3$, as $m_\infty$ increases, the mass of the non-uniform HP solution decreases. For sufficiently large $m_\infty$ (i.e. $L^{-1}\ll m_\infty\ll l_s^{-1}$), the non-uniform solution becomes localized in $\mathbb{S}_z^1$ and can be approximated by the spherically symmetric HP solution in four-dimensional noncompact space. However, as discovered above, the mass of the four-dimensional HP solution increases with $m_\infty$. This suggests that as $m_\infty$ continues to increase, the mass of the non-uniform solution will eventually stop decreasing and begin to increase. This scenario is consistent with the conjecture in section \ref{secd34} that the HP solutions, along with the black hole branch, form a swallowtail-type phase diagram. 

For $d=4$, since the mass varies slightly with $m_\infty$ as shown in (\ref{Cm}), the blue dot in figure \ref{smd4} can be replaced by a short curve. In particular, when the uniform solution has a mass below the critical value, $m_\infty$ is smaller than $m_{\infty,c}$. Thus, the size of the solutions—determined by $m_\infty^{-1}$—is larger than that at the critical point. This implies that the solution is stable.

The stability of the uniform solution below the critical mass can be checked by comparing its entropy with that of the non-uniform solution. Specifically, for the non-uniform solution, the subleading term in the entropy (\ref{SMFz}) is negative and scales as $1/G_N^{(d+2)}L\sim m_\infty/G_N^{(d+2)}$ near the critical point. In contrast, this term vanishes for the uniform HP solution~\cite{Chen:2021dsw}. Furthermore, when the quartic terms in the effective action are included, it follows from (\ref{exp}) that the subleading term in the entropy (\ref{SHP}) becomes proportional to $m_\infty^2/G_N^{(d+2)}L\sim m_\infty^3/G_N^{(d+2)}$, which is much smaller than the (absolute value of) corresponding term for the non-uniform solution, given that $\alpha'm^2_\infty\ll 1$. Therefore, the uniform solution indeed has a higher entropy than the non-uniform solution, indicating its stability against non-uniformity.

At the critical mass, rather than transitioning into a non-uniform HP solution, the uniform solution in $\mathbb{R}^4\times\mathbb{S}^1$ would transition into a localized black hole, which has a much higher entropy~\cite{Chu:2024ggi,Emparan:2024mbp}. As discussed in section \ref{secd34}, the black hole branch is expected to be connected with the non-uniform HP branch at $m_\infty\sim l_s^{-1}$. Consequently, in the microcanonical ensemble, the black hole, uniform string star and non-uniform HP branches together form a swallowtail-type phase diagram\footnote{Note that the uniform branch spans only a small range of $M$ around the critical value, as discussed above. However, when $\alpha'm_\infty^2$ is sufficiently small, quantum effects become significant, and the string star becomes a state of free strings, which extends to lower mass values~\cite{Chen:2021dsw}.}, similar to the one in the $d=3$ case.
\subsection{$d=5$}
In $\mathbb{R}^5\times \mathbb{S}_z^1$, a highly localized string star approaches the spherically symmetric solution in $\mathbb{R}^6$. As discussed before, the original HP EFT breaks down in $\mathbb{R}^6$, and the quartic terms have to be taken into account, giving the effective action (\ref{Id1}), with the higher-order terms still negligible. Note that these quartic terms break the rescaling property associated with (\ref{resHP}). Thus, the hatted variables and fields are no longer useful.

From the action (\ref{Id1}), we derive the e.o.m. as
 \begin{equation}
 \label{eomHPz4}
	\begin{split}
 	\nabla_x^2\chi+\partial_z^2\chi&=\left(m_{\infty}^2+\frac{\kappa}{\alpha'}\varphi+\frac{\kappa}{\alpha'}\varphi^2+\frac{\kappa}{2
    \alpha'}\chi^2\right)\chi \ ,\\
 	\nabla_x^2\varphi+\partial_z^2\varphi &=\frac{\kappa}{2\alpha'}(1+2\varphi)\chi^2\ .
	\end{split}
	\end{equation}
As in the previous section, we can use a localized solution as a seed for the relaxation method to generate non-uniform solutions. Interestingly, the localized solution can be approximated by~\cite{Balthazar:2022hno}
\begin{equation}
\label{d6HP}
    \chi(r,z)=-\sqrt{2}\varphi(r,z)=\sqrt{\frac{140\alpha'}{3\kappa}}\frac{m_\infty}{\left(1+\frac{1}{24}\sqrt{\frac{70\kappa}{3\alpha'}}m_\infty (r^2+z^2)\right)^2}\ .
\end{equation}

This analytic expression suggests that, in the numerical computation, it is convenient to rescale\footnote{Note that it is different from the previous rescaling (\ref{resHP}).} 
\begin{equation}
\label{res2}
\begin{split}
    r\to \frac{r}{\sqrt{m_\infty}}\ ,&\quad z\to \frac{z}{\sqrt{m_\infty}}\ ,\\
    \chi\to m_\infty\chi\ ,&\quad \varphi\to m_\infty\varphi\ .
\end{split}
\end{equation}
Otherwise, the field amplitudes and their variations would be so small that higher numerical resolution would be required.

For any solution of (\ref{eomHPz4}), substituting this e.o.m. into the on-shell action yields the following expression for the free energy.
\begin{equation}
\label{Fd5}
    F=\frac{\omega_{d-1}\kappa}{32\pi \alpha' G_N^{(d+2)}}\tilde{F}\ ,
\end{equation}
where
\begin{equation}
   \tilde{F}\equiv \int_0^{\frac{L}{2}}dz\int_0^\infty dr\ r^{d-1}\left(-2\varphi-4\varphi^2-\chi^2\right)\chi^2\ .
\end{equation}
Besides, using the second line of the e.o.m. (\ref{eomHPz4}), the mass can be computed similarly to (\ref{MC}) and is given by
\begin{equation}
\label{Md5}
    M=\frac{\omega_{d-1}}{8\pi G_N^{(d+2)}}\frac{\kappa}{\alpha'}\tilde{M}\ ,
\end{equation}
with
\begin{equation}
    \tilde{M}\equiv \int_0^{\frac{L}{2}}dz\int_0^\infty dr\ r^{d-1}\left(1+2\varphi(r,z)\right)\chi^2(r,z)\ .
\end{equation}

From these expressions for $F$ and $M$, the entropy can be calculated using (\ref{SMF}) as follows,
\begin{equation}
\begin{split}
    S=&\beta_HM+ \frac{\omega_{d-1}\kappa\beta_H}{32\pi \alpha' G_N^{(d+2)}}\tilde{S}\ .
\end{split}
\end{equation}
Here, we have defined
\begin{equation}
    \tilde{S}\equiv \frac{4\alpha'm_\infty^2}{\kappa} \tilde{M}-\tilde{F}\ .
\end{equation}
\begin{figure}
	\centering
	\subfigure[]{
	\begin{minipage}[t]{0.45\linewidth}
	\centering
	\includegraphics[width=2.5in]{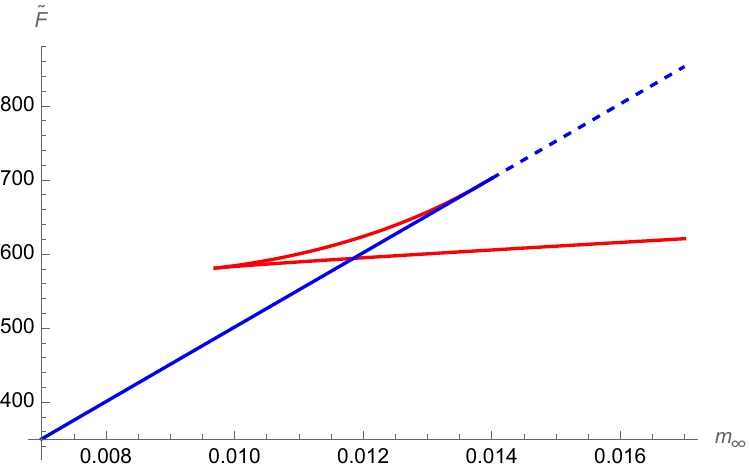}\label{fminfd5}
	\end{minipage}}
	\subfigure[]{
	\begin{minipage}[t]{0.45\linewidth}
	\centering
	\includegraphics[width=2.5in]{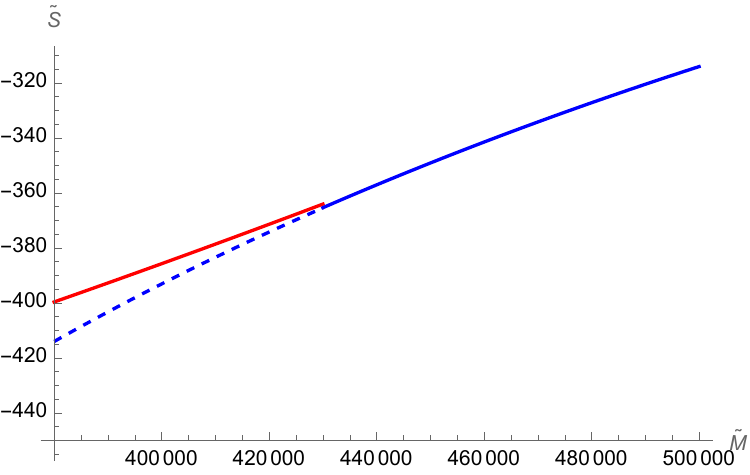}\label{smd5}
	\end{minipage}}
	\centering
\caption{\label{pdd5}Phase diagrams for the string stars at $d=5$ in (a) the canonical ensemble and (b) the microcanonical ensemble, for $L=200$ with $\frac{\kappa}{\alpha'}=1$. The red curves represent the non-uniform solutions. The blue curves denote the uniform solutions.}
\end{figure}

In figure \ref{pdd5}, we plot the phase diagrams for $L=200$, with $\kappa/\alpha'=1$. In the canonical ensemble, as shown in figure \ref{fminfd5}, the non-uniform solution exhibits a turning point as $m_\infty$ varies. This behavior contrasts with the cases of $d=3$ and $d=4$. Specifically, starting from a highly localized string star, as $m_\infty$ decreases, we find non-uniform solutions forming the lower red branch in figure \ref{fminfd5}. However, our numerical code fails to produce any solution at $m_\infty=0.0096$. This implies that solutions with greater uniformity exist at larger values of $m_\infty$.

To find these more uniform solutions, we create a seed for the relaxation method by linearly combining lower-branch solutions at $m_\infty=0.0097$ and $m_\infty=0.0098$,
\begin{equation}
    2(\chi,\varphi)|_{m_\infty=0.0097,\text{ lower branch}}-(\chi,\varphi)|_{m_\infty=0.0098,\text{ lower branch}}\ .
\end{equation}
Using this seed, we can generate the upper-branch solution at $m_\infty=0.0098$. Gradually increasing $m_\infty$, we then construct the upper red branch, which extends up to $m_\infty=0.014$, where the solution becomes uniform.

As shown in figure \ref{fminfd5}, the uniform and non-uniform solutions form a swallowtail-type phase diagram. In particular, as $m_\infty$ increases, the uniform HP solution undergoes a first-order phase transition at the critical point $m_\infty\approx 0.014$ (for $L=200$ with $\kappa/\alpha'=1$) and becomes non-uniform. Conversely, if we begin with a non-uniform solution at sufficiently large $m_\infty$ (but still much smaller than $l_s^{-1}$) and decrease $m_\infty$, the solution transitions into the uniform phase at the turning point $m_\infty\approx 0.0097$.

Note that perturbed non-uniform solutions are found at $m_\infty$ values smaller than the critical one, consistent with analyses that neglect the quartic terms in the action~\cite{Chu:2024ggi,Emparan:2024mbp}. However, as the non-uniformity grows, the quartic corrections gradually become significant, causing the non-uniform branch to turn around and extend towards larger values of $m_\infty$.

Finally, in the microcanonical ensemble, starting with a uniform solution at large mass, figure \ref{smd5} shows that as $M$ decreases, a second-order phase transition occurs at the critical point, resulting in a non-uniform solution. Note that the order of phase transition is different from that in the cases of $d=3$ and $d=4$. This finding also aligns with~\cite{Chu:2024ggi,Emparan:2024mbp}.

\subsection{$d=6$}\label{secd6}
We now consider the case of $d=6$, where both the HP effective action (\ref{Sphichiz}) and the quartic-corrected effective action (\ref{Id1}) become invalid for localized solutions. In this situation, we begin with the uniform solution at a small value of $m_\infty$, denoted by $\chi_*(r)$ and $\varphi_*(r)$, for which the quartic-corrected action (\ref{Id1}) is still applicable. By introducing non-uniform perturbations to the uniform solution, we aim to shed light on higher-dimensional solutions, which are not yet well understood.\footnote{See a recent paper~\cite{Bedroya:2024igb} for a non-perturbative prescription for the existence of higher-dimensional string star solutions. Note that this paper also leverages the non-uniformity of the HP solutions for $d\le 5$ in the presence of $\mathbb{T}^{n\ge 2}$ to explore higher-dimensional string star solutions. However, it is intriguing to ask whether the uniform solution transitions at the critical point directly into a string star localized in all directions of $\mathbb{T}^{n\ge 2}$, or only in one direction while remaining uniform along the others. It is this question that motivated the investigation of non-uniform string stars for $d=6$ in this subsection.}

First, we perturb the uniform solutions as follows,
\begin{equation}
\label{per}
    \chi(r,z)=\chi_*(r)+\lambda \cos \left(\frac{2\pi}{L}z \right)\chi_1(r)\ ,\quad \varphi(r,z)=\varphi_*(r)+\lambda \cos \left(\frac{2\pi}{L}z \right)\varphi_1(r) \ ,
\end{equation}
where $\lambda$ is the perturbative parameter. Plugging the perturbed fields in the e.o.m. (\ref{eomHPz4}), we derive the differential equations governing $\chi_1$ and $\varphi_1$. Specifically,
 \begin{equation}
 \label{pereomHPz4}
	\begin{split}
 	\nabla^2\chi_1-\left(\frac{2\pi}{L} \right)^2\chi_1&=m_{\infty}^2\chi_1+\frac{\kappa}{\alpha'}\left((\varphi_*+\varphi^2_*+\frac{3}{2}\chi^2_*)\chi_1+(\chi_*+2\chi_*\varphi_*)\varphi_1\right) \ ,\\
 	\nabla^2\varphi_1-\left(\frac{2\pi}{L}\right)^2\varphi_1 &=\frac{\kappa}{\alpha'}\left((\chi_*+2\chi_*\varphi_*)\chi_1+\chi_*^2\varphi_1\right)\ .
	\end{split}
	\end{equation}

As before, we can use the shooting method to solve the ordinary differential equations (\ref{pereomHPz4}), which yields $\chi_1$ and $\varphi_1$. Since these differential equations are linear, we can, for instance, set $\chi_1(r)=1$. Note that solving these equations amounts to solving an eigenvalue problem, as only a specific value of $L$ yields normalizable solutions. Accordingly, the shooting method requires iteratively adjusting $L$ until  normalizable solutions are obtained.

Once the perturbed solutions (\ref{per}) are obtained, they can be used as a seed for the relaxation method to generate non-uniform solutions.\footnote{This method of generating non-uniform solutions is utilized in~\cite{Emparan:2024mbp} also for $d=3$, 4 and 5.} For numerical efficiency, we again employ the rescaled variables and fields given in (\ref{res2}).

\begin{figure}
	\centering
	\subfigure[]{
	\begin{minipage}[t]{0.45\linewidth}
	\centering
	\includegraphics[width=2.5in]{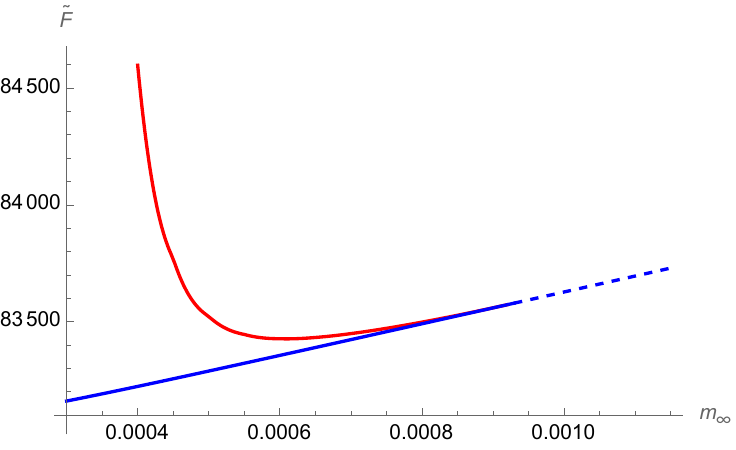}\label{Fminfd6per}
	\end{minipage}}
	\subfigure[]{
	\begin{minipage}[t]{0.45\linewidth}
	\centering
	\includegraphics[width=2.5in]{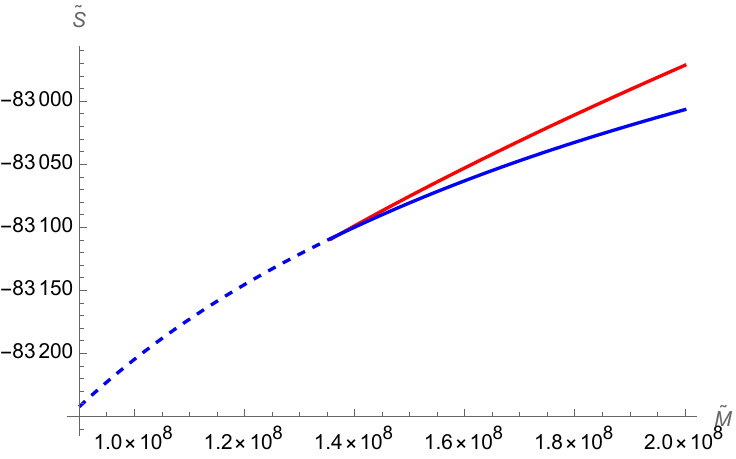}\label{SMd6per}
	\end{minipage}}
	\centering
\caption{\label{pdd6per}Phase diagrams for $d=6$ in the (a) canonical ensemble and (b) microcanonical ensemble, with $\frac{\kappa}{\alpha'}=1$ and $L=180$, near the critical point. The red curves represent the non-uniform solutions, while the blue curves denote the uniform solutions.}
\end{figure}
In figure \ref{pdd6per}, we plot the free energy and entropy for $L=180$ and $\kappa/\alpha'=1$ near the critical point $m_\infty\approx 0.00093$. As shown in figure \ref{Fminfd6per}, the perturbed non-uniform solutions are located at values of $m_\infty$ smaller than the critical one, where the uniform solutions (represented by the blue solid curve) have lower free energy and are therefore stable. This indicates that the uniform solution undergoes a first-order phase transition at the critical point. 

For the microcanonical ensemble (figure \ref{SMd6per}), the behavior is quite abnormal. In particular, the uniform solutions that are stable in the canonical ensemble (represented by the solid blue curve) become unstable, as they exhibit lower entropy than the non-uniform solutions. On the other hand, the uniform solutions with $m_\infty$ greater than the critical value (represented by the dashed blue curve) become stable in the microcanonical ensemble. This also implies that the uniform solution undergoes a second-order phase transition into a non-uniform solution at the critical mass, as the mass increases.

\begin{figure}
	\centering
	\subfigure[]{
	\begin{minipage}[t]{0.3\linewidth}
	\centering
	\includegraphics[width=1.8in]{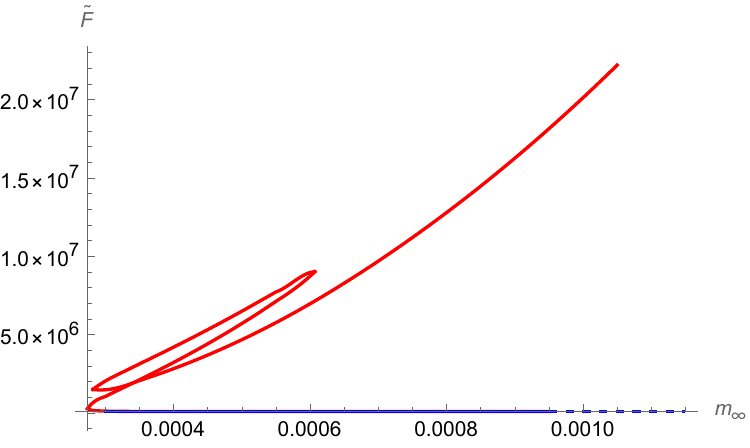}\label{Fminfd6}
	\end{minipage}}
	\subfigure[]{
	\begin{minipage}[t]{0.3\linewidth}
	\centering
	\includegraphics[width=1.8in]{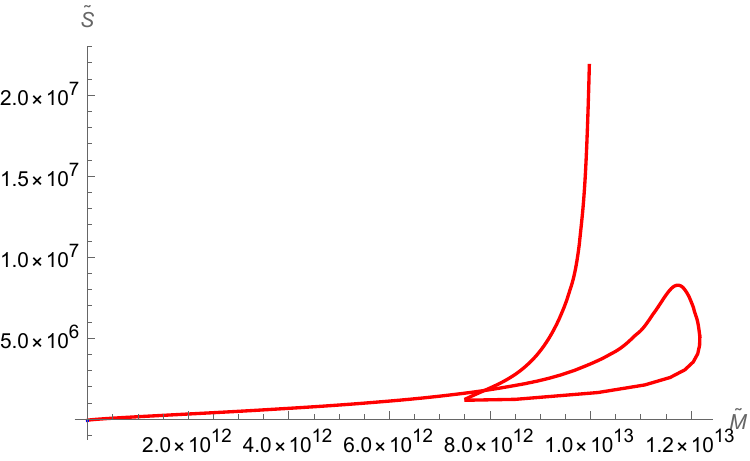}\label{SMd6}
	\end{minipage}}
        \subfigure[]{
	\begin{minipage}[t]{0.3\linewidth}
	\centering
	\includegraphics[width=1.8in]{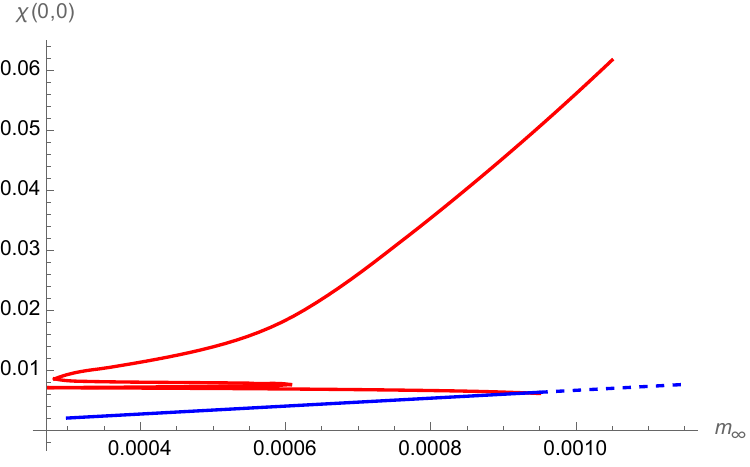}\label{chi0minfd6}
	\end{minipage}}
	\centering
\caption{\label{pdd6}Phase diagrams for $d=6$: (a) canonical ensemble, (b) microcanonical ensemble, and (c) $\chi(0,0)$ as a function of $m_\infty$, with $\frac{\kappa}{\alpha'}=1$ and $L=180$.}
\end{figure}
Zooming out from the scale of figure \ref{pdd6per}, we obtain the phase diagrams presented in figure \ref{pdd6}, which also includes the amplitude of the solution $\chi$. This additional figure reveals that the field amplitudes increase significantly at large $m_\infty$ (but still much smaller than $l_s^{-1}$), signaling the break-down of the effective action (\ref{Id1}).
\begin{figure}
	\centering
	\subfigure[]{
	\begin{minipage}[t]{0.45\linewidth}
	\centering
	\includegraphics[width=2.5in]{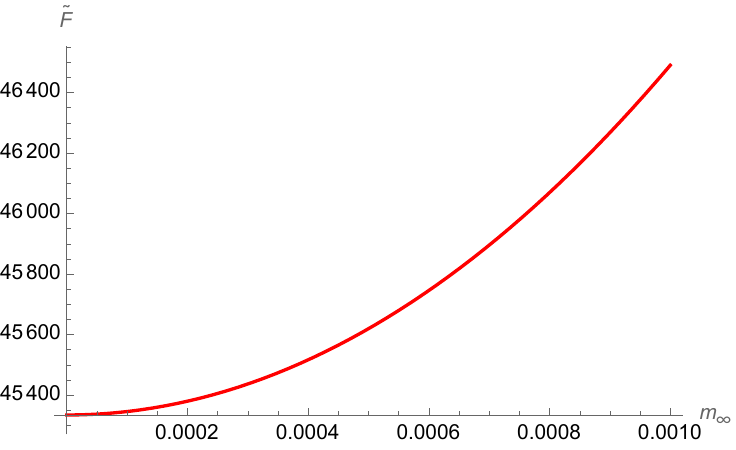}\label{Fminfd6loc}
	\end{minipage}}
        \subfigure[]{
	\begin{minipage}[t]{0.45\linewidth}
	\centering
	\includegraphics[width=2.5in]{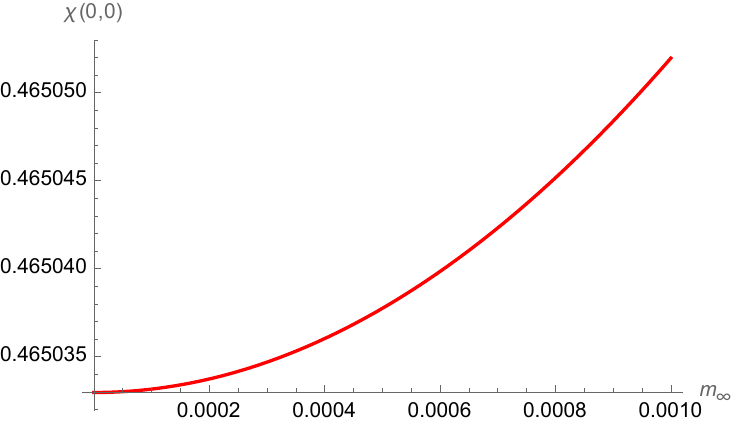}\label{chi0minfd6loc}
	\end{minipage}}
	\centering
\caption{\label{pdd6loc}Free energy and field amplitude of the localized solution for $d=6$, with $\frac{\kappa}{\alpha'}=1$ and $L=180$.}
\end{figure}

Also note that in the canonical ensemble, as seen in figure \ref{Fminfd6}, the free energy of the non-uniform solution is always higher than that of the uniform solution. This suggests that the uniform phase may never transition into a non-uniform one. However, this is not the case.\footnote{I thank Alek Bedroya and David Wu for discussions on this point.} As shown in figure \ref{Fminfd6loc}, we find an additional non-uniform branch, which is distinct from the one in figure \ref{Fminfd6}, and has lower free energy than the uniform branch (shown in figure \ref{Fminfd6per}). Therefore, it is this new branch that the uniform solution should transition into at the critical point $m_\infty\approx 0.00093$.

Of course, this solution is already beyond the regime of validity. For instance, figure \ref{chi0minfd6loc} shows that the field amplitudes are not much smaller than 1. Note that this solution actually corresponds to a higher-dimensional string star localized in the $z$ circle, and extends all the way to $m_\infty=0$. However, the precise value of the free energy of the localized solution is not important. What matters is that its free energy is basically independent of $L$, while that of the uniform solution is proportional to $L$. Therefore, for sufficiently large $L$, the uniform solution certainly has a higher free energy than the localized one, justifying our conclusion above about the transition from the uniform to the localized string star.

\section{Discussion}
The aim of this paper was to investigate the phase transitions between uniform and non-uniform string stars in the space $\mathbb{R}^d\times \mathbb{S}_z^1$. By analyzing the Horowitz-Polchinski EFT, we successfully reproduced the phase diagrams for $d=3$ and $d=4$ presented in~\cite{Emparan:2024mbp}. In particular, the phase transition at the critical point is second-order in the canonical ensemble and first-order in the microcanonical ensemble. For $d=2$, where no standard HP solution exists, we found novel solutions in which the Euclidean time circle opens up at infinity. Interestingly, unlike the cases of $d=3$ and $d=4$, the uniform and non-uniform solutions are disconnected. As discussed in section \ref{secd2}, this absence of a critical length for $L$ can be attributed to the scaling symmetry of the uniform solution, (\ref{scsy}).

We also went beyond the HP EFT by including the quartic terms in the effective action. First, this approach revealed that the mass of the uniform string star for $d=4$ increases with $m_\infty$. Second, we obtained the phase diagrams for $d=5$ and $d=6$. Specifically, at the critical point towards non-uniformity, the uniform string star undergoes a first-order phase transition in the canonical ensemble and a second-order transition in the microcanonical ensemble. These phenomena contrast with those for $d=3$ and $d=4$.

One interesting question is about the Lorentzian interpretation of the solutions studied in this paper. For black holes, the instability of the uniform solutions towards non-uniformity in Lorentzian spacetime is known as the Gregory-Laflamme instability, as mentioned before. Naturally, one might ask whether a similar Gregory-Laflamme instability exists for string stars. However, the winding tachyon $\chi$ has no counterpart in Lorentzian signature because Lorentzian time is not periodic. This produces an obstruction to continuing the solutions to Lorentzian spacetime. Therefore, understanding the nature of string stars in Lorentzian spacetime remains an important open question (see, e.g., ~\cite{Damour:1999aw,Kutasov:2005rr,Chialva:2009pf,Giveon:2020xxh,Jafferis:2021ywg} for related explorations).

Relatedly, recall that for $d=6$, figure \ref{SMd6per} indicates that at a given mass, when the spatial circle $L$ is smaller than the critical size, the uniform solution is entropically disfavored and unstable to non-uniform perturbations. This behavior is anomalous because, in Lorentzian signature, the Gregory-Laflamme instability does not occur when $L$ is smaller than the critical size.\footnote{I thank Roberto Emparan for pointing this out to me.} Therefore, our findings suggest that either something beyond classical dynamics is required for a Lorentzian interpretation of the string star, if such an interpretation exists, or the classical dynamics cannot be captured by the Euclidean EFT studied in this paper. This issue is left to future work.
\section*{Acknowledgement}
I would like to express my gratitude to Alek Bedroya, Roberto Emparan, David Kutasov, Mikel Sanchez-Garitaonandia, Marija Toma\v{s}evi\'{c} and David Wu for lots of interesting discussions and useful comments on the draft.

\bibliographystyle{utphys}
\bibliography{paper}{}

\providecommand{\href}[2]{#2}\begingroup\raggedright\begin{thebibliography}{10}

\bibitem{CALLAN1989673}
C.~Callan, R.~Myers, and M.~Perry, ``Black holes in string theory,'' \href{http://dx.doi.org/https://doi.org/10.1016/0550-3213(89)90172-7}{{\em Nuclear Physics B} {\bfseries 311} no.~3, (1989) 673--698}. \url{https://www.sciencedirect.com/science/article/pii/0550321389901727}.

\bibitem{Chen:2021qrz}
Y.~Chen, ``{Revisiting $R^4$ higher curvature corrections to black holes},'' \href{http://arxiv.org/abs/2107.01533}{{\ttfamily arXiv:2107.01533 [hep-th]}}.

\bibitem{Bowick:1985af}
M.~J. Bowick, L.~Smolin, and L.~C.~R. Wijewardhana, ``{Role of String Excitations in the Last Stages of Black Hole Evaporation},'' \href{http://dx.doi.org/10.1103/PhysRevLett.56.424}{{\em Phys. Rev. Lett.} {\bfseries 56} (1986) 424}.

\bibitem{Susskind:1993ws}
L.~Susskind, ``{Some speculations about black hole entropy in string theory},'' \href{http://arxiv.org/abs/hep-th/9309145}{{\ttfamily arXiv:hep-th/9309145}}.

\bibitem{Horowitz:1996nw}
G.~T. Horowitz and J.~Polchinski, ``{A Correspondence principle for black holes and strings},'' \href{http://dx.doi.org/10.1103/PhysRevD.55.6189}{{\em Phys. Rev. D} {\bfseries 55} (1997) 6189--6197}, \href{http://arxiv.org/abs/hep-th/9612146}{{\ttfamily arXiv:hep-th/9612146}}.

\bibitem{Sen:1995in}
A.~Sen, ``{Extremal black holes and elementary string states},'' \href{http://dx.doi.org/10.1142/S0217732395002234}{{\em Mod. Phys. Lett. A} {\bfseries 10} (1995) 2081--2094}, \href{http://arxiv.org/abs/hep-th/9504147}{{\ttfamily arXiv:hep-th/9504147}}.

\bibitem{Damour:1999aw}
T.~Damour and G.~Veneziano, ``{Selfgravitating fundamental strings and black holes},'' \href{http://dx.doi.org/10.1016/S0550-3213(99)00596-9}{{\em Nucl. Phys. B} {\bfseries 568} (2000) 93--119}, \href{http://arxiv.org/abs/hep-th/9907030}{{\ttfamily arXiv:hep-th/9907030}}.

\bibitem{Khuri:1999ez}
R.~R. Khuri, ``{Selfgravitating strings and string / black hole correspondence},'' \href{http://dx.doi.org/10.1016/S0370-2693(99)01265-4}{{\em Phys. Lett. B} {\bfseries 470} (1999) 73--76}, \href{http://arxiv.org/abs/hep-th/9910122}{{\ttfamily arXiv:hep-th/9910122}}.

\bibitem{Kutasov:2005rr}
D.~Kutasov, ``{Accelerating branes and the string/black hole transition},'' \href{http://arxiv.org/abs/hep-th/0509170}{{\ttfamily arXiv:hep-th/0509170}}.

\bibitem{Giveon:2005jv}
A.~Giveon and D.~Kutasov, ``{The Charged black hole/string transition},'' \href{http://dx.doi.org/10.1088/1126-6708/2006/01/120}{{\em JHEP} {\bfseries 01} (2006) 120}, \href{http://arxiv.org/abs/hep-th/0510211}{{\ttfamily arXiv:hep-th/0510211}}.

\bibitem{Giveon:2006pr}
A.~Giveon and D.~Kutasov, ``{Fundamental strings and black holes},'' \href{http://dx.doi.org/10.1088/1126-6708/2007/01/071}{{\em JHEP} {\bfseries 01} (2007) 071}, \href{http://arxiv.org/abs/hep-th/0611062}{{\ttfamily arXiv:hep-th/0611062}}.

\bibitem{Horowitz:1997jc}
G.~T. Horowitz and J.~Polchinski, ``{Selfgravitating fundamental strings},'' \href{http://dx.doi.org/10.1103/PhysRevD.57.2557}{{\em Phys. Rev. D} {\bfseries 57} (1998) 2557--2563}, \href{http://arxiv.org/abs/hep-th/9707170}{{\ttfamily arXiv:hep-th/9707170}}.

\bibitem{Brustein:2021cza}
R.~Brustein and Y.~Zigdon, ``{Black hole entropy sourced by string winding condensate},'' \href{http://dx.doi.org/10.1007/JHEP10(2021)219}{{\em JHEP} {\bfseries 10} (2021) 219}, \href{http://arxiv.org/abs/2107.09001}{{\ttfamily arXiv:2107.09001 [hep-th]}}.

\bibitem{Chen:2021emg}
Y.~Chen and J.~Maldacena, ``{String scale black holes at large D},'' \href{http://dx.doi.org/10.1007/JHEP01(2022)095}{{\em JHEP} {\bfseries 01} (2022) 095}, \href{http://arxiv.org/abs/2106.02169}{{\ttfamily arXiv:2106.02169 [hep-th]}}.

\bibitem{Chen:2021dsw}
Y.~Chen, J.~Maldacena, and E.~Witten, ``{On the black hole/string transition},'' \href{http://dx.doi.org/10.1007/JHEP01(2023)103}{{\em JHEP} {\bfseries 01} (2023) 103}, \href{http://arxiv.org/abs/2109.08563}{{\ttfamily arXiv:2109.08563 [hep-th]}}.

\bibitem{Urbach:2022xzw}
E.~Y. Urbach, ``{String stars in anti de Sitter space},'' \href{http://dx.doi.org/10.1007/JHEP04(2022)072}{{\em JHEP} {\bfseries 04} (2022) 072}, \href{http://arxiv.org/abs/2202.06966}{{\ttfamily arXiv:2202.06966 [hep-th]}}.

\bibitem{Balthazar:2022szl}
B.~Balthazar, J.~Chu, and D.~Kutasov, ``{Winding Tachyons and Stringy Black Holes},'' \href{http://arxiv.org/abs/2204.00012}{{\ttfamily arXiv:2204.00012 [hep-th]}}.

\bibitem{Balthazar:2022hno}
B.~Balthazar, J.~Chu, and D.~Kutasov, ``{On small black holes in string theory},'' \href{http://dx.doi.org/10.1007/JHEP03(2024)116}{{\em JHEP} {\bfseries 03} (2024) 116}, \href{http://arxiv.org/abs/2210.12033}{{\ttfamily arXiv:2210.12033 [hep-th]}}.

\bibitem{Bedroya:2022twb}
A.~Bedroya, ``{High energy scattering and string/black hole transition},'' \href{http://arxiv.org/abs/2211.17162}{{\ttfamily arXiv:2211.17162 [hep-th]}}.

\bibitem{Urbach:2023npi}
E.~Y. Urbach, ``{The black hole/string transition in AdS$_{3}$ and confining backgrounds},'' \href{http://dx.doi.org/10.1007/JHEP09(2023)156}{{\em JHEP} {\bfseries 09} (2023) 156}, \href{http://arxiv.org/abs/2303.09567}{{\ttfamily arXiv:2303.09567 [hep-th]}}.

\bibitem{Ceplak:2023afb}
N.~\v{C}eplak, R.~Emparan, A.~Puhm, and M.~Toma\v{s}evi\'c, ``{The correspondence between rotating black holes and fundamental strings},'' \href{http://dx.doi.org/10.1007/JHEP11(2023)226}{{\em JHEP} {\bfseries 11} (2023) 226}, \href{http://arxiv.org/abs/2307.03573}{{\ttfamily arXiv:2307.03573 [hep-th]}}.

\bibitem{Halder:2023nlp}
I.~Halder and D.~L. Jafferis, ``{Double winding condensate CFT},'' \href{http://dx.doi.org/10.1007/JHEP05(2024)189}{{\em JHEP} {\bfseries 05} (2024) 189}, \href{http://arxiv.org/abs/2308.11702}{{\ttfamily arXiv:2308.11702 [hep-th]}}.

\bibitem{Agia:2023skp}
N.~Agia and D.~L. Jafferis, ``{AdS$_3$ String Stars at Pure NSNS Flux},'' \href{http://arxiv.org/abs/2311.04956}{{\ttfamily arXiv:2311.04956 [hep-th]}}.

\bibitem{Bedroya:2024uva}
A.~Bedroya, C.~Vafa, and D.~H. Wu, ``{The Tale of Three Scales: the Planck, the Species, and the Black Hole Scales},'' \href{http://arxiv.org/abs/2403.18005}{{\ttfamily arXiv:2403.18005 [hep-th]}}.

\bibitem{Santos:2024ycg}
J.~E. Santos and Y.~Zigdon, ``{Self gravitating spinning string condensates},'' \href{http://dx.doi.org/10.1007/JHEP07(2024)217}{{\em JHEP} {\bfseries 07} (2024) 217}, \href{http://arxiv.org/abs/2403.20332}{{\ttfamily arXiv:2403.20332 [hep-th]}}.

\bibitem{Chu:2024ggi}
J.~Chu, ``{From black strings to fundamental strings: non-uniformity and phase transitions},'' \href{http://dx.doi.org/10.1007/JHEP04(2025)045}{{\em JHEP} {\bfseries 04} (2025) 045}, \href{http://arxiv.org/abs/2410.23597}{{\ttfamily arXiv:2410.23597 [hep-th]}}.

\bibitem{Emparan:2024mbp}
R.~Emparan, M.~Sanchez-Garitaonandia, and M.~Toma\v{s}evi\'c, ``{String theory in a pinch: resolving the Gregory-Laflamme singularity},'' \href{http://dx.doi.org/10.1007/JHEP02(2025)104}{{\em JHEP} {\bfseries 02} (2025) 104}, \href{http://arxiv.org/abs/2411.14998}{{\ttfamily arXiv:2411.14998 [hep-th]}}.

\bibitem{Ceplak:2024dxm}
N.~\v{C}eplak, R.~Emparan, A.~Puhm, and M.~Toma\v{s}evi\'c, ``{Size and Shape of Rotating Strings and the Correspondence to Black Holes},'' \href{http://arxiv.org/abs/2411.18690}{{\ttfamily arXiv:2411.18690 [hep-th]}}.

\bibitem{Bedroya:2024igb}
A.~Bedroya and D.~Wu, ``{String stars in $d\geq 7$},'' \href{http://arxiv.org/abs/2412.19888}{{\ttfamily arXiv:2412.19888 [hep-th]}}.

\bibitem{Montero:2022prj}
M.~Montero, C.~Vafa, and I.~Valenzuela, ``{The dark dimension and the Swampland},'' \href{http://dx.doi.org/10.1007/JHEP02(2023)022}{{\em JHEP} {\bfseries 02} (2023) 022}, \href{http://arxiv.org/abs/2205.12293}{{\ttfamily arXiv:2205.12293 [hep-th]}}.

\bibitem{Gubser:2001ac}
S.~S. Gubser, ``{On nonuniform black branes},'' \href{http://dx.doi.org/10.1088/0264-9381/19/19/303}{{\em Class. Quant. Grav.} {\bfseries 19} (2002) 4825--4844}, \href{http://arxiv.org/abs/hep-th/0110193}{{\ttfamily arXiv:hep-th/0110193}}.

\bibitem{Wiseman:2002zc}
T.~Wiseman, ``{Static axisymmetric vacuum solutions and nonuniform black strings},'' \href{http://dx.doi.org/10.1088/0264-9381/20/6/308}{{\em Class. Quant. Grav.} {\bfseries 20} (2003) 1137--1176}, \href{http://arxiv.org/abs/hep-th/0209051}{{\ttfamily arXiv:hep-th/0209051}}.

\bibitem{Kol:2002xz}
B.~Kol, ``{Topology change in general relativity, and the black hole black string transition},'' \href{http://dx.doi.org/10.1088/1126-6708/2005/10/049}{{\em JHEP} {\bfseries 10} (2005) 049}, \href{http://arxiv.org/abs/hep-th/0206220}{{\ttfamily arXiv:hep-th/0206220}}.

\bibitem{Sorkin:2004qq}
E.~Sorkin, ``{A Critical dimension in the black string phase transition},'' \href{http://dx.doi.org/10.1103/PhysRevLett.93.031601}{{\em Phys. Rev. Lett.} {\bfseries 93} (2004) 031601}, \href{http://arxiv.org/abs/hep-th/0402216}{{\ttfamily arXiv:hep-th/0402216}}.

\bibitem{Kol:2004pn}
B.~Kol and E.~Sorkin, ``{On black-brane instability in an arbitrary dimension},'' \href{http://dx.doi.org/10.1088/0264-9381/21/21/003}{{\em Class. Quant. Grav.} {\bfseries 21} (2004) 4793--4804}, \href{http://arxiv.org/abs/gr-qc/0407058}{{\ttfamily arXiv:gr-qc/0407058}}.

\bibitem{Kudoh:2005hf}
H.~Kudoh and U.~Miyamoto, ``{On non-uniform smeared black branes},'' \href{http://dx.doi.org/10.1088/0264-9381/22/19/004}{{\em Class. Quant. Grav.} {\bfseries 22} (2005) 3853--3874}, \href{http://arxiv.org/abs/hep-th/0506019}{{\ttfamily arXiv:hep-th/0506019}}.

\bibitem{Kol:2006vu}
B.~Kol and E.~Sorkin, ``{LG (Landau-Ginzburg) in GL (Gregory-Laflamme)},'' \href{http://dx.doi.org/10.1088/0264-9381/23/14/002}{{\em Class. Quant. Grav.} {\bfseries 23} (2006) 4563--4592}, \href{http://arxiv.org/abs/hep-th/0604015}{{\ttfamily arXiv:hep-th/0604015}}.

\bibitem{Harmark:2007md}
T.~Harmark, V.~Niarchos, and N.~A. Obers, ``{Instabilities of black strings and branes},'' \href{http://dx.doi.org/10.1088/0264-9381/24/8/R01}{{\em Class. Quant. Grav.} {\bfseries 24} (2007) R1--R90}, \href{http://arxiv.org/abs/hep-th/0701022}{{\ttfamily arXiv:hep-th/0701022}}.

\bibitem{Asnin:2007rw}
V.~Asnin, D.~Gorbonos, S.~Hadar, B.~Kol, M.~Levi, and U.~Miyamoto, ``{High and Low Dimensions in The Black Hole Negative Mode},'' \href{http://dx.doi.org/10.1088/0264-9381/24/22/015}{{\em Class. Quant. Grav.} {\bfseries 24} (2007) 5527--5540}, \href{http://arxiv.org/abs/0706.1555}{{\ttfamily arXiv:0706.1555 [hep-th]}}.

\bibitem{Emparan:2015gva}
R.~Emparan, R.~Suzuki, and K.~Tanabe, ``{Evolution and End Point of the Black String Instability: Large D Solution},'' \href{http://dx.doi.org/10.1103/PhysRevLett.115.091102}{{\em Phys. Rev. Lett.} {\bfseries 115} no.~9, (2015) 091102}, \href{http://arxiv.org/abs/1506.06772}{{\ttfamily arXiv:1506.06772 [hep-th]}}.

\bibitem{Emparan:2018bmi}
R.~Emparan, R.~Luna, M.~Mart\'\i{}nez, R.~Suzuki, and K.~Tanabe, ``{Phases and Stability of Non-Uniform Black Strings},'' \href{http://dx.doi.org/10.1007/JHEP05(2018)104}{{\em JHEP} {\bfseries 05} (2018) 104}, \href{http://arxiv.org/abs/1802.08191}{{\ttfamily arXiv:1802.08191 [hep-th]}}.

\bibitem{Figueras:2022zkg}
P.~Figueras, T.~Fran\c{c}a, C.~Gu, and T.~Andrade, ``{Endpoint of the Gregory-Laflamme instability of black strings revisited},'' \href{http://dx.doi.org/10.1103/PhysRevD.107.044028}{{\em Phys. Rev. D} {\bfseries 107} no.~4, (2023) 044028}, \href{http://arxiv.org/abs/2210.13501}{{\ttfamily arXiv:2210.13501 [hep-th]}}.

\bibitem{Gregory:1993vy}
R.~Gregory and R.~Laflamme, ``{Black strings and p-branes are unstable},'' \href{http://dx.doi.org/10.1103/PhysRevLett.70.2837}{{\em Phys. Rev. Lett.} {\bfseries 70} (1993) 2837--2840}, \href{http://arxiv.org/abs/hep-th/9301052}{{\ttfamily arXiv:hep-th/9301052}}.

\bibitem{Gregory:1994bj}
R.~Gregory and R.~Laflamme, ``{The Instability of charged black strings and p-branes},'' \href{http://dx.doi.org/10.1016/0550-3213(94)90206-2}{{\em Nucl. Phys. B} {\bfseries 428} (1994) 399--434}, \href{http://arxiv.org/abs/hep-th/9404071}{{\ttfamily arXiv:hep-th/9404071}}.

\bibitem{Dias:2015nua}
O.~J.~C. Dias, J.~E. Santos, and B.~Way, ``{Numerical Methods for Finding Stationary Gravitational Solutions},'' \href{http://dx.doi.org/10.1088/0264-9381/33/13/133001}{{\em Class. Quant. Grav.} {\bfseries 33} no.~13, (2016) 133001}, \href{http://arxiv.org/abs/1510.02804}{{\ttfamily arXiv:1510.02804 [hep-th]}}.

\bibitem{Bogojevic:1990hv}
A.~R. Bogojevic and L.~Perivolaropoulos, ``{Black holes in a periodic universe},'' \href{http://dx.doi.org/10.1142/S021773239100035X}{{\em Mod. Phys. Lett. A} {\bfseries 6} (1991) 369--376}.

\bibitem{Frolov:2003kd}
A.~V. Frolov and V.~P. Frolov, ``{Black holes in a compactified space-time},'' \href{http://dx.doi.org/10.1103/PhysRevD.67.124025}{{\em Phys. Rev. D} {\bfseries 67} (2003) 124025}, \href{http://arxiv.org/abs/hep-th/0302085}{{\ttfamily arXiv:hep-th/0302085}}.

\bibitem{Brustein:2021ifl}
R.~Brustein and Y.~Zigdon, ``{Effective field theory for closed strings near the Hagedorn temperature},'' \href{http://dx.doi.org/10.1007/JHEP04(2021)107}{{\em JHEP} {\bfseries 04} (2021) 107}, \href{http://arxiv.org/abs/2101.07836}{{\ttfamily arXiv:2101.07836 [hep-th]}}.

\bibitem{Chialva:2009pf}
D.~Chialva, ``{Self-interacting fundamental strings and black holes},'' \href{http://dx.doi.org/10.1016/j.nuclphysb.2009.04.026}{{\em Nucl. Phys. B} {\bfseries 819} (2009) 256--281}, \href{http://arxiv.org/abs/0903.3977}{{\ttfamily arXiv:0903.3977 [hep-th]}}.

\bibitem{Giveon:2020xxh}
A.~Giveon, N.~Itzhaki, and U.~Peleg, ``{Instant Folded Strings and Black Fivebranes},'' \href{http://dx.doi.org/10.1007/JHEP08(2020)020}{{\em JHEP} {\bfseries 08} (2020) 020}, \href{http://arxiv.org/abs/2004.06143}{{\ttfamily arXiv:2004.06143 [hep-th]}}.

\bibitem{Jafferis:2021ywg}
D.~L. Jafferis and E.~Schneider, ``{Stringy ER = EPR},'' \href{http://dx.doi.org/10.1007/JHEP10(2022)195}{{\em JHEP} {\bfseries 10} (2022) 195}, \href{http://arxiv.org/abs/2104.07233}{{\ttfamily arXiv:2104.07233 [hep-th]}}.

\end{thebibliography}\endgroup

\end{document}